\documentclass[11pt,a4paper]{article}
\pdfoutput=1
\usepackage{jcappub}
\usepackage{mathtools}
\usepackage{bm} 
\usepackage{xcolor}
\RequirePackage{epsfig}

\newcommand{\be}{\begin{equation}}
\newcommand{\ee}{\end{equation}}
\newcommand{\ba}{\begin{eqnarray}}
\newcommand{\ea}{\end{eqnarray}}
\newcommand{\bi}{\begin{itemize}}
\newcommand{\ei}{\end{itemize}}

\newcommand{\bfi}{\begin{figure}
\epsfxsize=9cm
\epsffile}
\newcommand{\bfinew}{\begin{figure}
\begin{center}
\includegraphics}
\newcommand{\efi}{\end{figure}}
\newcommand{\efinew}{
\end{center}
\end{figure}}
\newcommand{\no}{\nonumber}

\graphicspath{ {./figures/} }

\usepackage{graphicx}
\usepackage{epstopdf}
\usepackage{subfigure}

\DeclarePairedDelimiter\ang{\langle}{\rangle}
\DeclarePairedDelimiter\bang{\Big\langle}{\Big\rangle}

\title{Polarized Sunyaev Zel'dovich tomography}

\author[a]{Anne-Sylvie Deutsch}
\author[b,c]{Matthew C. Johnson}
\author[c]{Moritz M{\"u}nchmeyer}
\author[b,c]{Alexandra Terrana}
\preprint{IGC-17/5-1}

\affiliation[a]{Institute for Gravitation and the Cosmos and Physics Department, The Pennsylvania State University, University Park, PA 16802, USA}
\affiliation[b]{Department of Physics and Astronomy, York University, Toronto, Ontario, M3J 1P3, Canada}
\affiliation[c]{Perimeter Institute for Theoretical Physics, Waterloo, Ontario N2L 2Y5, Canada}

\emailAdd{asdeutsch@psu.edu}
\emailAdd{aterrana@perimeterinstitute.ca}
\emailAdd{mjohnson@perimeterinstitute.ca}
\emailAdd{mmunchmeyer@perimeterinstitute.ca}

\abstract{Secondary CMB polarization is induced by the late-time scattering of CMB photons by free electrons on our past light cone. This \emph{polarized Sunyaev Zel'dovich} (pSZ) effect is sensitive to the electrons' locally observed CMB quadrupole, which is sourced primarily by long wavelength inhomogeneities. 
By combining the remote quadrupoles measured by free electrons throughout the Universe after reionization, the pSZ effect allows us to obtain additional information about large scale modes beyond what can be learned from our own last scattering surface. Here we determine the power of pSZ tomography, in which the pSZ effect is cross-correlated with the density field binned at several redshifts, to provide information about the long wavelength Universe. The signal we explore here is a power asymmetry in the cross-correlation between $E$ or $B$ mode CMB polarization and the density field. We compare this to the cosmic variance limited noise: the random chance to get a power asymmetry in the absence of a large scale quadrupole field. By computing the necessary transfer functions and cross-correlations, we compute the signal-to-noise ratio attainable by idealized next generation CMB experiments and galaxy surveys. We find that a signal-to-noise ratio of $\sim 1-10$ is in principle attainable over a significant range of power multipoles, with the strongest signal coming from the first multipoles in the lowest redshift bins. These results prompt further assessment of realistically measuring the pSZ signal and the potential impact for constraining cosmology on large scales.}

\begin{document}
\maketitle
\flushbottom

\section{Introduction}

The direct observation of the primary cosmic microwave background (CMB) anisotropies has driven the era of precision cosmology, casting light on the contents and evolution of the Universe, and providing compelling evidence for the standard cosmological model, $\Lambda$CDM. The hunt for beyond-the-standard-cosmological-model (BSCM) physics has been ongoing, with the WMAP~\cite{Bennett2010} and Planck~\cite{PlanckCollaboration2015} CMB satellites providing a few tantalizing clues for BSCM physics on the largest observable scales. These include a lack of correlations on large scales, the low CMB quadrupole power, the alignment of the temperature quadrupole and octupole, the Cold Spot, a hemispherical power asymmetry, and other anomalies (for a recent review see ref.~\cite{Schwarz:2015cma}). Unfortunately, the statistical significance of such large scale anomalies cannot be assessed further using temperature anisotropies alone  because the statistical and systematic errors in existing measurements are dominated by cosmic variance~\footnote{If the large-angle CMB polarization can be faithfully extracted from foregrounds, there will be some progress. }. 

However, there is another promising avenue to explore in the coming era of precision measurements of the secondary CMB, which is dominated by lensing and the Sunyaev Zel'dovich effect~\footnote{Here, we consider only the nearly frequency independent contributions, assuming that strongly frequency independent contributions such as the thermal Sunyaev Zel'dovich effect and the cosmic infrared background have been perfectly removed.}. The scattering of photons onto our light-cone allows us to access additional realizations of the longest modes probed by the CMB, as the scattered photons originate from the bulk, and therefore carry information about modes inside our light-cone. Extracting this information could decrease the cosmic variance error on the largest modes, possibly alleviating (or confirming!) large scale anomalies. Two promising avenues include CMB lensing, which can be used to reconstruct our fundamental dipole~\cite{Meerburg:2017xga}, and the kinetic Sunyaev Zel'dovich effect~\cite{Zhang10d,Zhang11b,Zhang:2015uta,Terrana2016}. However, in this paper, we focus on the ability of the polarized Sunyaev-Zel'dovich (pSZ) effect~\cite{Sunyaev1980,Sazonov1999,Audit1999,Challinor2000,Itoh2000,Emritte2016,2012PhRvD..85l3540A}, the induced CMB polarization due to Thomson scattering from free electrons after reionization, to provide constraints on large scale modes. 

Thomson scattering of photons on a free electron creates a linear polarization pattern that depends on the quadrupole observed by the electron. Therefore, CMB polarization could be used to reconstruct distant quadrupoles, possibly yielding information about the large scale Universe. This idea, first proposed in ref.~\cite{Kamionkowski1997}, has received significant attention in the literature. Previous work has explored the detectability of this effect based on the observation of CMB polarization in the direction of galaxy clusters~\cite{Sazonov1999,Hall2014,Challinor2000}, suggesting that the pSZ effect could be a target for the next generation of CMB experiments. The degree to which the pSZ effect gives constraints on primordial, large scale modes has been explored in refs.~\cite{Seto:2000uc,Abramo:2006gp,Bunn2006,Liu2016,Maartens2011}. One important point is that the local quadrupole for different clusters are correlated with each other~\cite{Portsmouth2004}, making it important to assess the degree to which cosmic variance is in fact alleviated~\cite{Bunn2006}. Because tensors contribute to the quadrupole, the pSZ effect has been proposed as a probe of primordial gravitational waves~\cite{2012PhRvD..85l3540A}. Finally, remote quadrupole measurements could have other implications for determining the properties of dark energy~\cite{Baumann2003,Seto2005} and the intra-cluster medium~\cite{Lavaux2004}.

In this paper, we lay the theoretical foundation for pSZ tomography, the cross-correlation of CMB polarization arising from the pSZ effect with probes of large scale structure at different redshifts. This cross-correlation has an intrinsically statistically anisotropic component -- a power asymmetry -- which encodes the variation of the locally observed quadrupole along our past light cone. Performing this cross-correlation at different redshifts allows one to obtain three-dimensional information about the structure of the Universe on very large scales. More precisely, the contribution to the CMB polarization from the pSZ effect is proportional to $q_{\rm eff}(1+\delta_e)$, where $q_{\rm eff}$ is the remote quadrupole field, and $\delta_e$ is the electron density field. The cross-correlation with $\delta_e$ at different redshifts has a statistically anisotropic component $\langle (Q \pm i U) \delta_e \rangle \sim q_{\rm eff} \langle \delta_e \delta_e\rangle$, which is a long-wavelength modulation of small-scale power. In this paper, we characterize this signal by exploiting a decomposition of the Stokes parameters into curl-free $E$ modes and curl $B$-modes and performing a multipolar decomposition of the cross-power as a function of redshift.

Measurements of the CMB polarization in the direction of galaxy clusters can be thought of as a special case of pSZ tomography, which focuses on the largest amplitude pixels in a pSZ map. However, the distribution of free electrons is a continuous field, implying that in principle there is more information to gather (this point was also highlighted in ref.~\cite{2012PhRvD..85l3540A}). One main result of this paper is to determine how much information there is to gather about the remote quadrupole field from the power asymmetry in the cosmic variance limit. Another result is to forecast the requirements for next generation of CMB experiments and galaxy surveys to measure the remote quadrupole field. An important further step is to connect the information that can be gathered from the remote quadrupole field to CMB anomalies and constraints on cosmological parameters. We pursue this question in future work.

The paper is organized as follows. Section~\ref{sec:pSZeffect} is devoted to the pSZ effect, where we derive the pSZ signal and its power spectrum. In section~\ref{sec:pSZtomo}, we cross-correlate the pSZ signal with tracers of electron density at known redshift in order to find the pSZ tomography signal and its cosmic variance. In section~\ref{sec:detection} we discuss the detectability of the signal. We present our conclusions in section~\ref{sec:conclusion}. A number of calculations are collected in a set of appendices.

\section{The polarized SZ effect}
\label{sec:pSZeffect}

\begin{figure}[htbp]
	\centering
	\includegraphics[width=8cm]{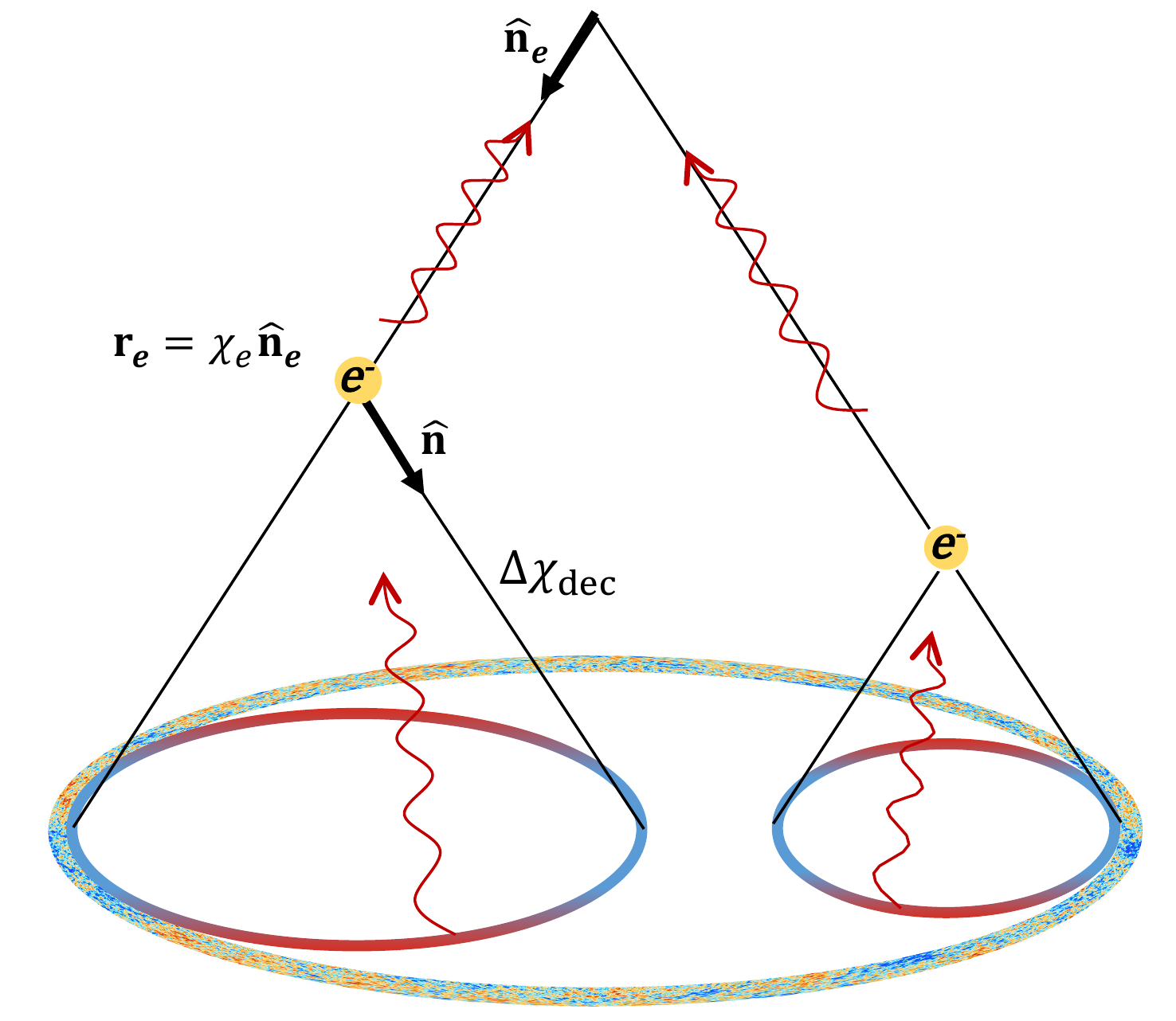}
	\caption{CMB photons scattering off free electrons on our past light cone. An electron's position is given by ${\bf r}_e = \chi_e{\bf \hat{n}}_e$. From the electron's location, the direction to a point on its surface of last scattering is ${\bf \hat{n} }$ and the distance to last scattering is $\Delta \chi_\text{dec}$. Notice that the scattering provides information about CMB photons from inside our past light cone. Specifically, the pSZ effect is sensitive to the locally observed quadrupole.}
	\label{fig:lightcone}
\end{figure}

Unpolarized light incident on free electrons becomes polarized due to Thomson scattering. The polarization generated depends on the quadrupole observed by the free electron. The polarized SZ (pSZ) effect can therefore be thought of as a census of the remote quadrupole as observed by free electrons on our past light cone. The contribution to the Stokes parameters $Q \pm i U$ is given by an integral along the line of sight as
\begin{equation}
(Q \pm iU)^{\text{pSZ}} ({\bf \hat{n}}_e) = - \frac{\sqrt{6}}{10}  \int d\chi_e \frac{d \tau ({\bf \hat{n}}_e,\chi_e)}{d\chi_e} \ e^{-\tau ({\bf \hat{n}}_e,\chi_e)} \sum_{m=-2}^{2} q^{m}_{\rm eff} ({\bf \hat{n}}_e, \chi_e) \left._{\pm 2}Y_{2 m}\right. ({\bf \hat{n}}_e) .
\label{eq:effective-quadrupole-to-QiU}
\end{equation}
Here, the optical depth $\tau$ is defined by,
\begin{equation}\label{eq:opticaldepth}
\tau ({\bf \hat{n}}_e,\chi_e) = \int_{0}^{\chi_e} d \chi \ \sigma_T a_e n_e ({\bf \hat{n}}_e,\chi), \ \ \  \frac{d \tau ({\bf \hat{n}}_e,\chi_e)}{d\chi_e} = \sigma_T a_e n_e ({\bf \hat{n}}_e,\chi_e),
\end{equation}
where $\sigma_T$ is the Thomson cross-section, $n_e({\bf \hat{n}}_e, \chi_e)$ is the electron number density, ${\bf \hat{n}}_e$ denotes the angular direction to the free electron on the sky, and $\chi_e$ is the comoving radial coordinate to the electron along our past light cone,
\begin{equation}
\chi_e =  \int_0^{z_e} \frac{d z}{H(z)} =  -\int_{1}^{a_e} \ \frac{da}{H(a)a^2} ,
\end{equation}
where $z_e$ and $a_e$ are the electron's redshift and scale factor respectively. Figure~\ref{fig:lightcone} describes the geometry.

Well after reionization, we can approximate by substituting eq.~\eqref{eq:opticaldepth} into eq.~\eqref{eq:effective-quadrupole-to-QiU}, obtaining
\begin{align}
  (Q \pm iU)^{\text{pSZ}} ({\bf \hat{n}}_e) = & - \frac{\sqrt{6} \ \sigma_T}{10}  \int d\chi_e \  a_e \bar{n}_e (\chi_e) (1+\delta_e ({\bf \hat{n}}_e,\chi_e)) \\
  & \qquad \times \sum_{m=-2}^{2} q^{m}_{\rm eff} ({\bf \hat{n}}_e, \chi_e) \left._{\pm 2}Y_{2 m}\right. ({\bf \hat{n}}_e), 
  \label{eq:QiUpSZ}
\end{align}
where we have written the electron number density as $n_e({\bf \hat{n}}_e, \chi_e) = \bar{n}_e(\chi_e)(1+\delta_e({\bf \hat{n}}_e, \chi_e))$ in terms of the average electron number density $\bar{n}_e(\chi_e)$, and the density contrast $\delta_e$. From here on, we will assume that electrons trace dark matter, and therefore we use $\delta$ rather than $\delta_e$. We make this assumption for simplicity, and this will not change the main results presented below. We defer a more accurate model for the electron distribution to future work.

The key quantity of interest, denoted by $q^m_{\rm eff} ({\bf \hat{n}}_e, \chi_e)$, is the CMB quadrupole observed by an electron along our past light cone in the ${\bf \hat{n}}_e$ direction at comoving distance $\chi_e$ (or alternatively, located at ${\bf r}_e = \chi_e {\bf \hat{n}}_e$):
\begin{equation}
q^{m}_{\rm eff} ({\bf \hat{n}}_e, \chi_e) = \int_\Omega d^2 {\bf \hat{n}} \ \Theta(\chi_e, {\bf \hat{n}}_e, {\bf \hat{n}}) \ Y^*_{2 m} ({\bf \hat{n}}).
\label{eq:effective-quadrupole-def}
\end{equation}
There are three contributions to the local CMB temperature at the position of the electron: the Sachs-Wolfe effect (SW) due to gravitational redshifting at the surface of last scattering, the integrated Sachs-Wolfe effect (ISW) resulting from the late-time evolution of gravitational potential, and the Doppler effect from the relative motion of electrons at the surface of last scattering and at an observer's location. We can write each contribution in terms of the primordial Newtonian gravitational potential $\Psi_i$:
\begin{align}
	\Theta _{\rm SW} ({\bf \hat{n}}_e, \chi_e, {\bf \hat{n}}) = & \left( 2D_\Psi(\chi_\text{dec}) -\frac{3}{2} \right) \Psi_i({\bf r}_\text{dec}), \label{eqn:SW} \\
	\Theta _{\rm ISW} ({\bf \hat{n}}_e, \chi_e, {\bf \hat{n}}) = & \ 2\int_{a_{\rm
    dec}}^{a_e} \frac{dD_\Psi}{da}\Psi_i({\bf r}(a)) da, \label{eqn:ISW} \\
    	\Theta_{\rm Doppler} ({\bf \hat{n}}_e, \chi_e, {\bf \hat{n}}) = & \ {\bf \hat{n}}\cdot
\left[ D_v(\chi_\text{dec})\nabla\Psi_i({\bf r}_\text{dec})-D_v(\chi_e)\nabla\Psi_i({\bf r}_e) \right]. \label{eq:thetaDopp}
\end{align}
In the above equations, we define the distance as ${\bf r}(a) \equiv \chi_e {\bf \hat{n}}_e + \Delta \chi(a) {\bf \hat{n}}$ with $\Delta\chi(a) = -\int_{a_e}^{a} da' [H(a')a'^2]^{-1}$ (see figure~\ref{fig:lightcone}). Specifically, ${\bf r}_\text{dec}={\bf r}(a_\text{dec})$ and $\Delta\chi_\text{dec} = \Delta\chi(a_\text{dec})$, while the comoving distance to decoupling is simply $\chi_\text{dec}=-\int_1^{a_\text{dec}}da[H(a)a^2]^{-1}$. We have introduced the growth function, $D_\Psi(\chi)$, which relates the potential to its primordial value at $a \rightarrow 0$ through the definition 
\begin{equation}
\Psi({\bf r},\chi) =D_\Psi(\chi)\Psi_i({\bf r}).
\end{equation}
On superhorizon scales, we can employ the approximation
\begin{equation}
\label{eqn:PhiSH}
D_\Psi(a)\equiv \frac{\Psi_{\rm SH}(a)}{\Psi_{\rm
  SH,i}}=\frac{16\sqrt{1+y}+9y^3+2y^2-8y-16}{10y^3} \left[ \frac{5}{2}\Omega_m \frac{E(a)}{a}\int_0^a
\frac{da}{E^3(a) \ a^3} \right] ,
\end{equation}
where $y\equiv a/a_\text{eq}$ and $E(a)\equiv\sqrt{\Omega_ma^{-3}+\Omega_\Lambda}$ is the normalized Hubble parameter. The velocity growth function $D_v$ in eq.~\eqref{eq:thetaDopp}, defined by ${\bf v}({\bf r},\chi) = -D_v(\chi) \nabla\Psi_i({\bf r})$, is given in terms of scale factor as,
\begin{equation}
\label{eqn:Dv}
D_v(a)\equiv \frac{2a^2H(a)}{H^2_0\Omega_m} \frac{y}{4+3y}
\left[D_\Psi+\frac{dD_\Psi}{d\ln a}\right].
\end{equation}

\subsection{Fourier kernel for the effective quadrupole}
In this subsection, we relate the effective quadrupole $q^{m}_{\rm eff}$ given in eq.~\eqref{eq:effective-quadrupole-def} to the primordial gravitational potential, $\Psi_i$, through the expressions given in~\eqref{eqn:SW}-\eqref{eq:thetaDopp}. Firstly express $\Psi_i$ in Fourier space as

\begin{equation} \label{eq:psifourier}
\Psi_i ({\bf r}) = \int \frac{d^3 k}{(2 \pi)^3} \tilde{\Psi}_i ({\bf k}) e^{i \chi_e {\bf k} \cdot {\bf \hat{n}}_e } e^{i \Delta \chi {\bf k} \cdot {\bf \hat{n}} },
\end{equation}
where we have explicitly expanded the position ${\bf r} = \chi_e {\bf \hat{n}}_e + \Delta \chi {\bf \hat{n}}$. Inserting the expressions for each contribution to the CMB temperature~\eqref{eqn:SW}-\eqref{eq:thetaDopp} in \eqref{eq:effective-quadrupole-def}, we obtain (see Appendix~\ref{app:kernel} for the details of the calculation):
\begin{equation}
q^{m}_{\rm eff} ({\bf \hat{n}}_e, \chi_e) = \int \frac{d^3k}{(2 \pi)^3}  \tilde{\Psi}_i({\bf k}) T(k)  \left[ \mathcal{G}_{\rm SW} + \mathcal{G}_{\rm ISW} + \mathcal{G}_{\rm Doppler } \right] Y_{2 m}^*({\bf \hat{k}}) \ e^{i \chi_e {\bf k} \cdot {\bf \hat{n}}_e},
\label{eq:initial-density-to-effective-quadrupole}
\end{equation}
where the kernels $\mathcal{G}_{\rm SW}$, $\mathcal{G}_{\rm ISW}$ and $\mathcal{G}_{\rm Doppler }$ are given by:
\begin{align} 
	\mathcal{G}_{\rm SW}(k,\chi_e) = & -4\pi \left( 2D_\Psi(\chi_{\rm dec}) -\frac{3}{2} \right) j_{2} (k \Delta \chi_\text{dec}), \no \\
	\mathcal{G}_{\rm ISW}(k,\chi_e) = & -8\pi \int_{a_{\rm dec}}^{a_e} da \frac{dD_\Psi}{da}  \ j_{2} (k \Delta \chi (a)), \no \\
	\mathcal{G}_{\rm Doppler }(k,\chi_e) = & \frac{4\pi}{5} k D_v(\chi_\text{dec}) \left[ 3j_3(k\Delta\chi_\text{dec}) - 2j_1(k\Delta\chi_\text{dec}) \right]. \label{eq:kernel}
\end{align}
We have also incorporated the transfer function, $T(k)$, in the above expression which we approximate using the BBKS fitting function~\cite{Bardeen1986},
\begin{equation} \label{eq:Tk}
T(k) = \frac{\ln \left[ 1 + 0.171 x \right]}{0.171 x} \left[ 1+ 0.284 x + (1.18 x)^2 + (0.399 x)^3 + (0.49 x)^4 \right]^{-0.25}\ ,
\end{equation}
where $x = k / k_{\rm eq}$ with $k_{\rm eq} = a_{\rm eq} H(a_{\rm eq}) = \sqrt{2/a_{\rm eq}} H_0 \simeq 82.5 H_0$. 

\begin{figure}[htbp]
	\centering
	\subfigure{\includegraphics[width=.44\textwidth]{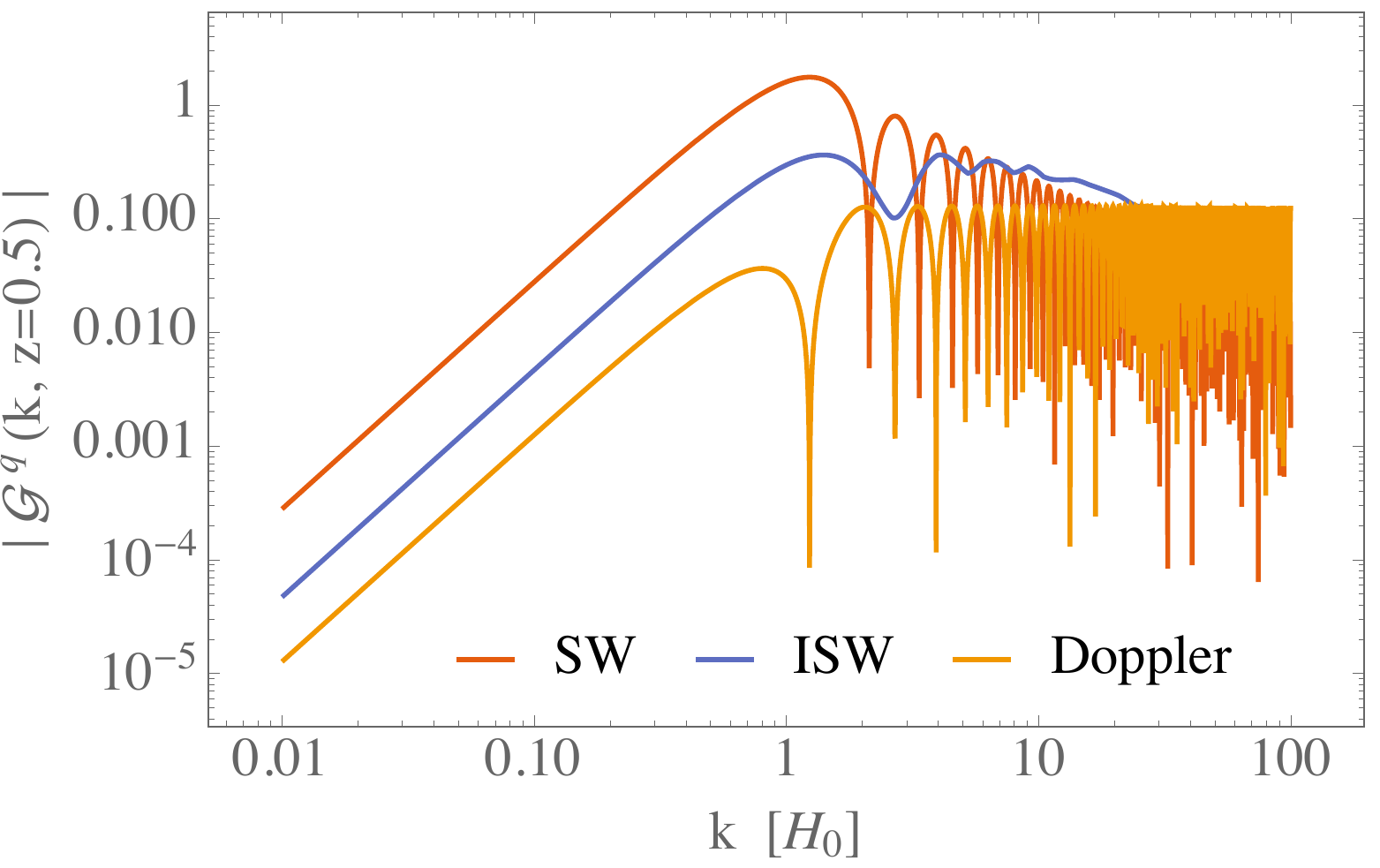}}
	\hspace{.05\textwidth}
	\subfigure{\includegraphics[width=.44\textwidth]{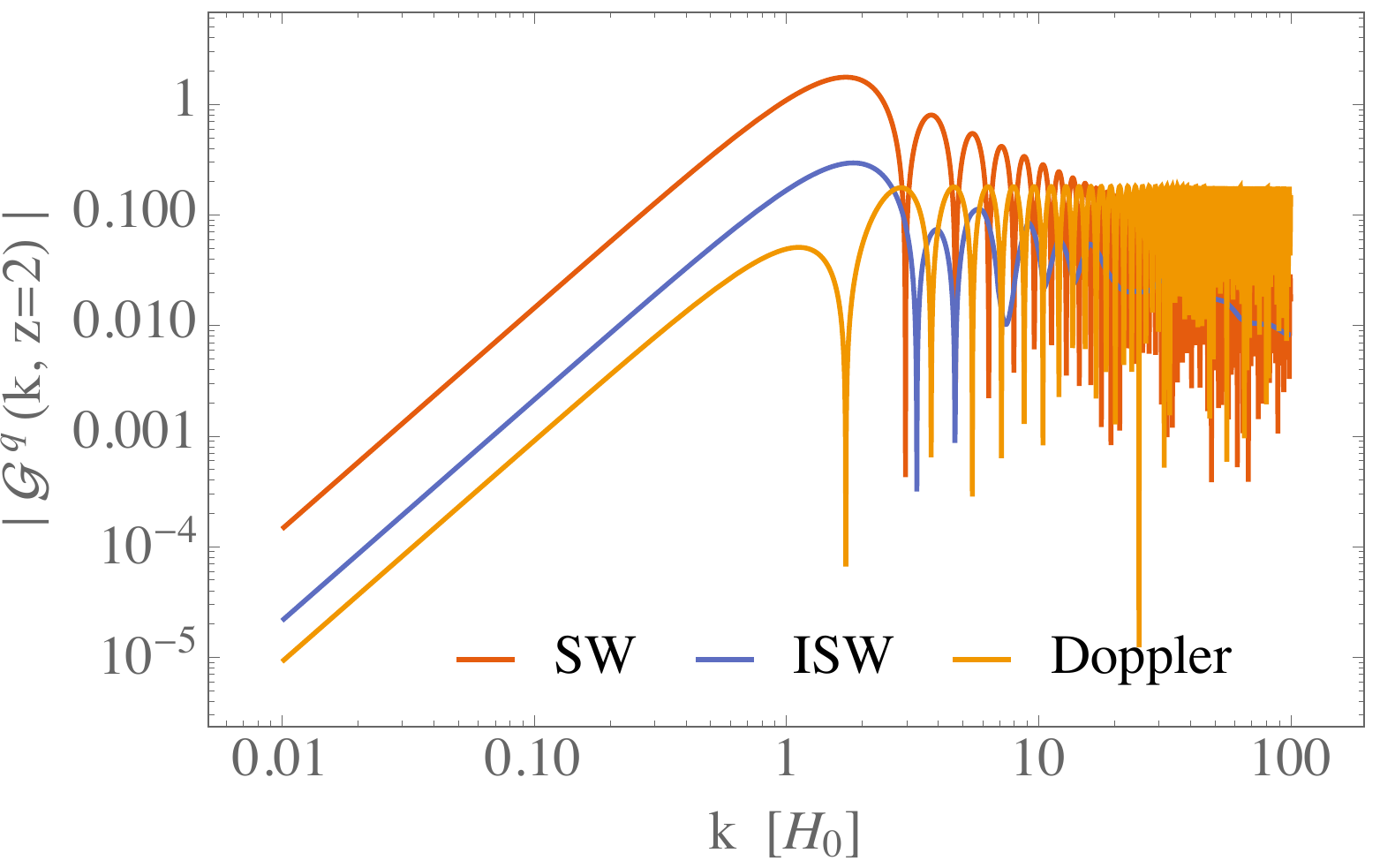}}
	\caption{The three contributions to the kernel, $\mathcal{G}_{\rm SW}(k,\chi_e)$, $\mathcal{G}_{\rm ISW}(k,\chi_e) $, and $\mathcal{G}_{\rm Doppler}(k,\chi_e)$ in eq.~\eqref{eq:kernel}. The left panel is evaluated at $z_e=0.5$, and the right is evaluated at $z_e=2$.}
	\label{fig:kernel}
\end{figure}

The kernels in eq.~\ref{eq:kernel} are plotted in figure~\ref{fig:kernel} at $z_e = 0.5$ (left) and $z_e = 2.0$ (right). The dominant contribution to the SW and ISW kernels are from large scales, with the amplitude peaking near $k \sim H_0$. On large scales, the SW and Doppler contributions are negative, while the ISW contribution is positive. To leading order in $k$, each kernel is proportional to $k^2$ in the limit $k \rightarrow 0$. Note that there is only a partial cancellation between the SW, Doppler and ISW contributions to the quadrupole field. On sufficiently small scales, $k \gtrsim 20 H_0$, the Doppler contribution dominates the SW and ISW components of the quadrupole field. Finally, comparing the left and right panel of figure~\ref{fig:kernel}, the kernel is more sensitive to larger scales at lower redshifts (as expected from the geometry in Figure~\ref{fig:lightcone}).

\subsection{Angular decomposition of the effective quadrupole}
It is convenient to define a total effective quadrupole that is the sum of the projections of $q^m_{\text{eff}}$ on the basis of spin-weighted spherical harmonics:
\begin{equation}\label{eq:qeffeff}
	\tilde{q}_\text{eff}^\pm({\bf \hat{n}}_e,\chi_e) \equiv \sum_{m=-2}^2 q^{m}_{\rm eff} ({\bf \hat{n}}_e, \chi_e) \left._{\pm 2}Y_{2 m}\right. ({\bf \hat{n}}_e) \ . 
\end{equation}
We also make use of the multipolar expansion of this quantity,
\begin{equation}\label{eq:qsumalm}
	\tilde{q}_\text{eff}^\pm({\bf \hat{n}}_e,\chi_e) =\sum_{\ell m}a_{\ell m}^q(\chi_e) \left._{\pm 2}Y_{\ell m}\right. ({\bf \hat{n}}_e) \ ,
\end{equation}
as well as the inverse relation to solve for the mulitpole coefficients,
\begin{equation}\label{eq:almqdef}
	a_{\ell m}^q(\chi_e) = \int_\Omega d^2{\bf \hat{n}}_e\ \tilde{q}_\text{eff}^\pm({\bf \hat{n}}_e,\chi_e) \ {}_{\pm 2}Y_{\ell m}^* ({\bf \hat{n}}_e) \ .
\end{equation}
Using the definition for $\tilde{q}_\text{eff}^\pm({\bf \hat{n}}_e,\chi_e)$, eq.~\eqref{eq:qeffeff}, and the result for $q^m_{\text{eff}}({\bf \hat{n}}_e,\chi_e)$, eq.~\eqref{eq:initial-density-to-effective-quadrupole}, we can relate $a_{\ell m}^q(\chi_e)$ to the primordial gravitational potential.
The step-by-step calculation, given in Appendix~\ref{app:almq}, results in the final expression:
\begin{equation} \label{eq:almqfinal}
	a_{\ell m}^q(\chi_e) = \int \frac{d^3k}{(2\pi)^3} \Delta_\ell^q(k,\chi_e)\ \tilde{\Psi}_i({\bf k})\ Y^*_{\ell m}({\bf \hat{k}}),
\end{equation}
where the transfer function for the quadrupole is 
\begin{equation}\label{eq:Deltaq}
	\Delta_\ell^q(k,\chi_e) = -5 i^\ell \sqrt{\frac{3}{8}}\sqrt{\frac{(\ell+2)!}{(\ell-2)!}} \frac{j_\ell(k\chi_e)}{(k\chi_e)^2}\ T(k)\ \mathcal{G}^{q}(k,\chi_e), 
\end{equation}
with $\mathcal{G}^{q}(k,\chi_e) \equiv \mathcal{G}_{\rm SW}(k,\chi_e) + \mathcal{G}_{\rm ISW}(k,\chi_e) + \mathcal{G}_{\rm Doppler }(k,\chi_e)$ from eq.~\eqref{eq:kernel}. The transfer function is zero for $\ell =0$ and $\ell =1$. Comparing eq.~\eqref{eq:qeffeff} and eq.~\eqref{eq:qsumalm}, note that if the quadrupole field was only a function of redshift $q^{m}_{\rm eff}  = q^{m}_{\rm eff} (\chi_e)$, we would have $a_{2 m}^q(\chi_e) = q^{m}_{\rm eff} (\chi_e)$. Therefore, $\ell = 2$ is probing the ``monopole" moment of the 5 independent components of the quadrupole field. Note further that at $\chi_e = 0$, the $q^{m}_{\rm eff}$ are simply the quadrupole moments of the CMB temperature anisotropies observed here on Earth.

\subsection{Polarized SZ power spectrum}
\label{sec:polarized-sz-power-spectrum}

Neglecting foregrounds, the measured polarization of the CMB arises from scattering of CMB photons near the time of decoupling and from the reionization and post-reionization era. The latter includes a contribution from the average density of free electrons which dominates at low-$\ell$ (the ``reionization bump"), as well as the pSZ effect, which arises from spatial variations in the electron density and dominates at high-$\ell$. We make use of the decomposition of the polarization anisotropies into a curl-free component ($E$-modes) and a curl component ($B$-modes)~\cite{Kamionkowski:1996zd}. For a homogeneous distribution of electrons, scalar contributions to the quadrupole source only $E$-modes and tensor contributions to the quadrupole source a combination of $E$-modes and $B$-modes. An inhomogeneous distribution of electrons, which is considered here, induces a $B$-mode even where there is no tensor contribution to the quadrupole field. This also occurs in CMB lensing (see e.g.~\cite{Hu:1997hv}) and patchy reionization~\cite{Dvorkin:2008tf,Dvorkin:2009ah}. In this section, we calculate the $E$-mode and $B$-mode power spectrum induced by the pSZ effect, and compare it with the lensed primary CMB $E$-modes and $B$-modes. We neglect tensor contributions to the quadrupole (see ref.~\cite{2012PhRvD..85l3540A}, which treats this case in detail), and assume a linear bias of order unity between the dark matter and electron distribution. 

$E$ and $B$ modes can be defined on the full sky in terms of the spin-2 harmonic expansion coefficients of $(Q\pm iU)({\bf \hat{n}}_e)$:
\begin{eqnarray}\label{eq:EandBdecomp}
E_{\ell m} &=& \frac{1}{2} \left( {}_{+ 2} a_{\ell m} + {}_{- 2} a_{\ell m} \right) ,\\
B_{\ell m} &=& \frac{1}{2i} \left( {}_{+ 2} a_{\ell m} - {}_{- 2} a_{\ell m} \right),
\end{eqnarray}
where
\begin{equation}
{}_{\pm 2} a_{\ell m} = \int d^2 \hat{n}_e \ (Q\pm iU)({\bf \hat{n}}_e) {}_{\pm 2}Y_{\ell m}({\bf \hat{n}}_e)^* .
\end{equation}
The $E$-mode contribution from the pSZ effect has two contributions:
\begin{eqnarray}
E_{\ell m}^{\rm pSZ} = E_{\ell m}^{\rm pSZ,(0)} +  E_{\ell m}^{\rm pSZ,(1)}.
\end{eqnarray}
$E_{\ell m}^{\rm pSZ,(0)}$ is the contribution from the homogeneous density of free electrons, given by
\begin{align}\label{eq:pSZE0}
E_{\ell m}^{\rm pSZ,(0)} =& - \frac{\sqrt{6} \ \sigma_T}{10}  \int d\chi_e \  a_e \bar{n}_e (\chi_e)  a_{\ell m}^q (\chi_e),
\end{align}
and $E_{\ell m}^{\rm pSZ,(1)}$ is the contribution from the variation in density, given by
\begin{align}\label{eq:pSZE1}
E_{\ell m}^{\rm pSZ,(1)} =& - \frac{\sqrt{6} \ \sigma_T}{10}  \int d\chi_e \  a_e \bar{n}_e (\chi_e) \sum_{\ell',m'}\sum_{\ell'',m''} \frac{1}{2} \left( 1 + (-1)^{\ell+\ell'+\ell''}\right) \sqrt{\frac{(2 \ell+1) (2\ell'+1) (2\ell''+1)}{4 \pi}} \no \\
& \times \left(\begin{array}{ccc} \ell & \ell'' & \ell' \\ -m & m'' & m' \end{array} \right) \left(\begin{array}{ccc} \ell & \ell'' & \ell' \\ -2 & 2 & 0 \end{array} \right) a_{\ell' m'}^\delta (\chi_e) a_{\ell'' m''}^q (\chi_e).
\end{align}
The $B$-mode contribution from pSZ arises only because of the variation in density, and is given by
\begin{align}\label{eq:pSZB1}
B_{\ell m}^{\rm pSZ,(1)} =& - \frac{\sqrt{6} \ \sigma_T}{10}  \int d\chi_e \  a_e \bar{n}_e (\chi_e) \sum_{\ell',m'}\sum_{\ell'',m''} \frac{1}{2} \left( 1 - (-1)^{\ell+\ell'+\ell''}\right) \sqrt{\frac{(2 \ell+1) (2\ell'+1) (2\ell''+1)}{4 \pi}} \no \\
& \times \left(\begin{array}{ccc} \ell & \ell'' & \ell' \\ -m & m'' & m' \end{array} \right) \left(\begin{array}{ccc} \ell & \ell'' & \ell' \\ -2 & 2 & 0 \end{array} \right) a_{\ell' m'}^\delta (\chi_e) a_{\ell'' m''}^q (\chi_e).
\end{align}

We define angular power spectra for the density and quadrupole fields on the basis of spin-weighted and normal spherical harmonics respectively:
\begin{align}\label{eq:qq_def}
	\bang{\tilde{q}_\text{eff}^\pm({\bf \hat{n}}_e,\chi_e) \tilde{q}_\text{eff}^\pm({\bf \hat{n}}'_e,\chi'_e)} = & \sum_{L,M} C_L^{qq}(\chi_e,\chi'_e)\ {}_{\pm 2}Y_{LM}({\bf \hat{n}}_e)  \ {}_{\pm 2}Y^*_{LM}({\bf \hat{n}}'_e), \\
	\bang{\delta ({\bf \hat{n}}_e,\chi_e)\delta ({\bf \hat{n}}'_e,\chi'_e)} = & \sum_{L',M'} C_{L'}^{\delta\delta}(\chi_e,\chi'_e)\ Y_{L'M'}({\bf \hat{n}}_e)\  Y^*_{L'M'}({\bf \hat{n}}'_e) \label{eq:angdelta},
\end{align}
where $C_L^{qq}(\chi_e,\chi'_e)$ and $C_{L'}^{\delta\delta}(\chi_e,\chi'_e)$ are given by
\begin{align}
	C_L^{qq}(\chi_e,\chi'_e) = & \langle  a_{L M}^q (\chi_e)^* a_{L M}^q (\chi_e') \rangle \no \\
	= & \int \frac{dk\ k^2}{(2\pi)^3} \ P_\Psi(k)\ \Delta_L^{q,*}(k,\chi_e)  \Delta_L^q(k,\chi'_e)  \label{eq:Clqq_chi}\\
	C_{L'}^{\delta\delta}(\chi_e,\chi'_e) = & \langle  a_{L M}^\delta (\chi_e)^* a_{L M}^\delta (\chi_e') \rangle \no \\
	= & \int \frac{dk\ k^2}{(2\pi)^3} \ 4\pi\ j_{L'}(k\chi_e) \sqrt{P_\delta(k,\chi_e)}\ 4\pi\ j_{L'}(k\chi'_e) \sqrt{P_\delta(k,\chi'_e)}, \label{eq:Cldeltadelta_chi}.
\end{align}
Here, $P_\Psi(k)$ is the power spectrum of the gravitational potential, satisfying $\ang{ \tilde{\Psi}_i({\bf k})\tilde{\Psi}_i({\bf k'})} =(2\pi)^3\delta^{(3)}({\bf k-k'})P_\Psi(k)$, and $P_\delta(k,\chi)$ is the non-linear matter power spectrum, which was computed using the Cosmicpy package.\footnote{See cosmicpy.github.io} 

The polarization power spectra for the pSZ effect involve the correlation function $\ang{(1+\delta)q(1+\delta')q'}$. Using our assumption of Gaussian fields with zero mean, we simplify this to $\ang{qq'}+\ang{qq'}\ang{\delta\delta'}$. We expect the cross term $\ang{q\delta'}\ang{q'\delta}$ to be negligibly small since $q$ and $\delta$ contribute on very different scales, so their respective transfer functions have little overlap in $\ell$. Focusing first on the $E$-mode power spectrum, we can again use our assumption of Gaussian fields to write
\begin{equation}
C_\ell^{EE,\text{pSZ}} =  C_\ell^{EE,\text{pSZ},(0)} + C_\ell^{EE,\text{pSZ},(1)} .
\end{equation}
The power spectrum for the first term is equal to (see Appendix~\ref{sec:contributions-from-qq} for the details of the calculation):
\begin{equation}
\begin{split}
	C_\ell^{EE,\text{pSZ}, (0)} & \simeq \frac{6\sigma_T^2}{100} \int \frac{d\chi}{\chi^2} P_\Psi(k) \ a_e^2(\chi)\ \bar{n}_e^2(\chi) \left(\frac{5}{4\pi} \sqrt{\frac{3}{8}}\sqrt{\frac{(\ell+2)!}{(\ell-2)!}}\right)^2 \\
	& \quad \times \left[ \frac{T(k)}{(k\chi)^2} \left[ \mathcal{G}_{\rm SW}(k,\chi) + \mathcal{G}_{\rm ISW}(k,\chi) +  \mathcal{G}_{\rm Doppler}(k,\chi) \right] \right]^2 \Bigg|_{k \rightarrow (\ell+1/2)/\chi},
\end{split}
\label{eq:ClEpSZqq}
\end{equation}
where the integral runs from $\chi=0$ to reionization. Note that this is simply the standard contribution to the $E$-mode polarization (the reionization bump) in the limit where reionization is instantaneous.

The power spectrum coming from the second term is (see Appendix~\ref{sec:contributions-from-qq-dd}):
\begin{equation}\label{eq:ClEpSZ1}
	C_\ell^{EE,\text{pSZ}, (1)} = \frac{6\sigma_T^2}{100} \sum_{L,L'} \frac{(2L+1)(2L'+1)}{4\pi}  \frac{1}{2}\left( 1 + (-1)^{\ell+L+L'}\right)  \left(\begin{array}{ccc} \ell & L & L' \\ \mp 2 & \pm 2 & 0 \end{array} \right)^2 f_{L,L'}
\end{equation}
where $f_{L,L'}$ is given by
\begin{equation}
	f_{L,L'} \simeq \int \frac{d\chi}{\chi^2} \ C_L^{qq}(\chi) \ a^2_e(\chi)\  \bar{n}^2_e(\chi)\ P_\delta\left( \frac{L' + 1/2}{\chi},\chi \right) .
\end{equation}
This is the contribution to the $E$-mode polarization arising due to variations in the small-scale distribution of free electrons.

The $B$-mode power spectrum is (see Appendix~\ref{sec:contributions-from-qq-dd}):
\begin{equation}\label{eq:ClBpSZ1}
	C_\ell^{BB,\text{pSZ}, (1)} = \frac{6\sigma_T^2}{100} \sum_{L,L'} \frac{(2L+1)(2L'+1)}{4\pi} \frac{1}{2}  \left(1 - (-1)^{\ell+L+L'}\right)  \left(\begin{array}{ccc} \ell & L & L' \\ \mp 2 & \pm 2 & 0 \end{array} \right)^2 f_{L,L'},
\end{equation}
with the same $f_{L,L'}$ as given above. Both $E$ and $B$-mode contributions to the pSZ power have the same behavior at high $\ell$.

The lensed $E$-mode and $B$-mode power spectra (assuming no primordial tensors) are computed using CAMB~\cite{Lewis:1999bs} at low-$\ell$, and extrapolated to high $\ell$ assuming that the dominant contribution to the $E$-mode power spectrum arises from lensing of the primary CMB $E$-modes. To estimate the lensing contribution to $C_\ell^{EE}$, we use the approximation from ref.~\cite{Lewis:2006aa}, valid at $\ell \gg 3000$,
\begin{equation} \label{eq:lensing_high_ell_approx}
	C_\ell^{EE,\text{lensed}} = \frac{1}{2} \ell^2 C_\ell^{\phi\phi} R^E,
\end{equation}
where $C_\ell^{\phi\phi}$ is the lensing potential (see Appendix \ref{sec:lensingpotential}) and $R^E$ is defined by
\begin{equation} \label{eq:RE}
	R^E = \frac{1}{4\pi} \int{d\ell \ \ell^3 \ C_\ell^{EE,\text{unlensed}}}\sim 2 \times 10^7 \mu\text{K}^2.
\end{equation}
At high-$\ell$, the lensing $B$-modes are equal to the lensing $E$-modes~\cite{Lewis:2006aa}, and so we set $C_\ell^{BB,\text{lensed}} = C_\ell^{EE,\text{lensed}}$.

\begin{figure}
	\centering
	\includegraphics[width=.7\textwidth]{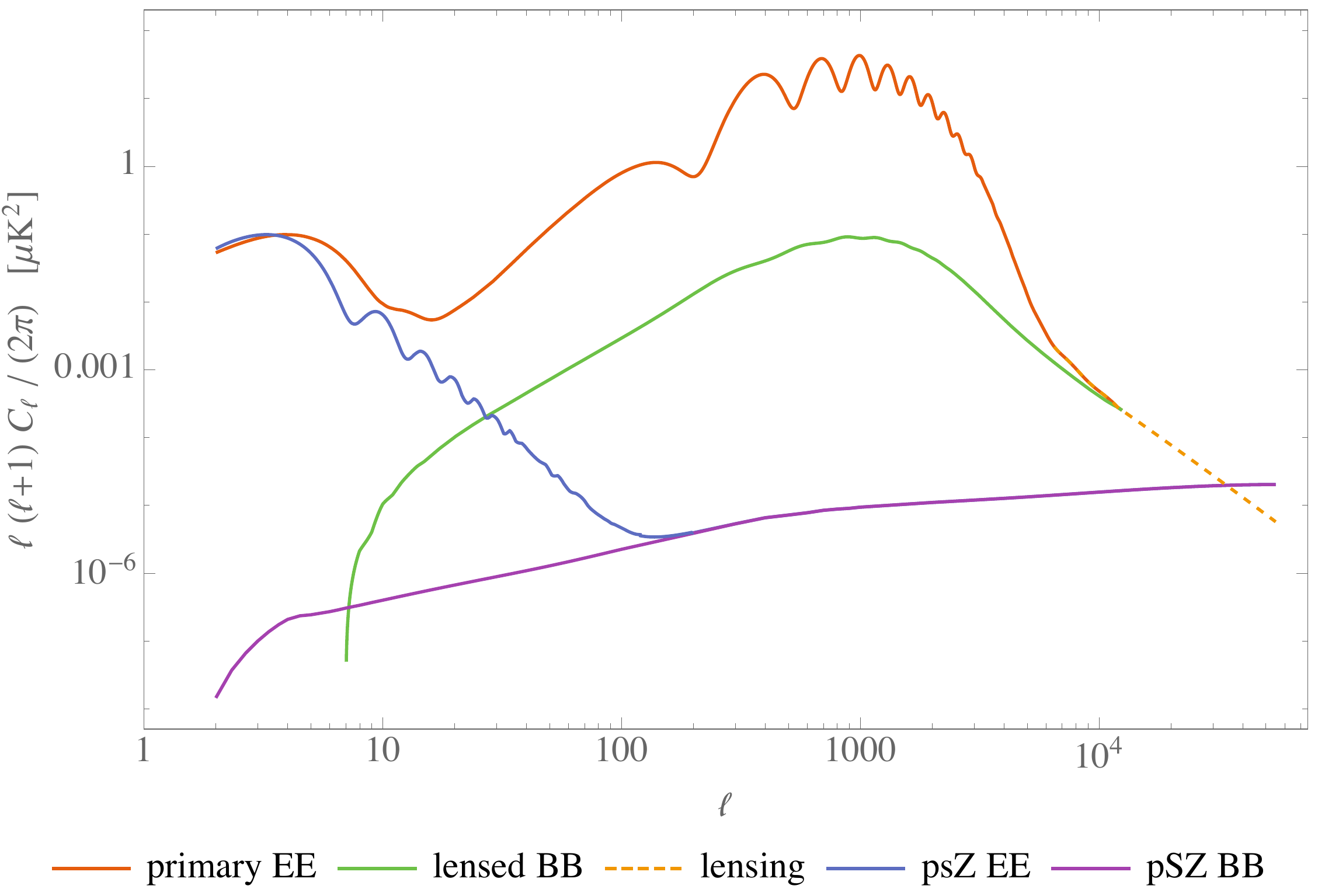}
	\caption{The polarized SZ contribution to the $E$-mode and $B$-mode power spectra in comparison to the primary lensed power spectra (assuming no primordial tensors). $C_\ell^{EE,\text{pSZ}, (0)}$ contributes only at low $\ell$, while $C_\ell^{EE,\text{pSZ}, (1)}$  and $C_\ell^{BB,\text{pSZ}, (1)}$ dominate the power spectrum around $\ell \gtrsim 3.4 \times 10^4$. The primary CMB from CAMB is extrapolated to high $\ell$ using the lensing approximation~\eqref{eq:lensing_high_ell_approx}.}
	\label{fig:ClEEpSZ}
\end{figure}

Figure~\ref{fig:ClEEpSZ} shows the contributions from $C_\ell^{EE,\text{pSZ}, (0)}$, $C_\ell^{EE,\text{pSZ}, (1)}$, and $C_\ell^{BB,\text{pSZ}, (1)}$ to the polarized SZ spectrum in comparison to the lensed primary $E$-mode and $B$-mode power spectra. $C_\ell^{EE,\text{pSZ}, (0)}$ gives a contribution to the largest scales of the power spectrum, and is in rough agreement with the result from CAMB (which treats reionization more consistently than we do here). On the other hand, $C_\ell^{EE,\text{pSZ}, (1)}$ and $C_\ell^{BB,\text{pSZ}, (1)}$ are much smaller, and become comparable to the primary CMB only at $\ell \gtrsim 3.4 \times 10^4$. 

\section{Polarized SZ tomography}
\label{sec:pSZtomo}

The pSZ signal is tiny, but it induces a statistically anisotropic cross-correlation with the matter distribution that can be detectable. We call this approach pSZ tomography, in analogy with kSZ tomography~\cite{Ho09,Shao11b, Zhang11b, Zhang01,Munshi:2015anr,2016PhRvD..93h2002S,Ferraro:2016ymw,Hill:2016dta,Terrana2016}. The pSZ signal depends on the local electron density as well as the local CMB quadrupole. In a direction $\hat{\bf n}_e$ and at a comoving distance $\chi_e$ the value of the quadrupole $q^\pm_{\rm eff}(\hat{\bf n}_e,\chi_e)$ modulates the strength of the cross-correlation between the CMB polarization and the electron density field. The CMB quadrupole is a slowly varying, large scale field, while the electron density has a lot of small scale structure. A probe of the quadrupole field is therefore given by the large scale modulation of the local correlation of the high $\ell$ CMB with the matter distribution. This data is the input in our tomography estimator. We now make these ideas precise and estimate the signal to noise.

\subsection{Correlation between polarisation and matter due to the pSZ}\label{eq:longshortsplitsec}

We first derive the cross correlation between polarisation and matter due to the pSZ effect. We assume the most ideal scenario, in which we have knowledge of the electron density field, which we further assume to trace the dark matter. We also assume a purely Gaussian primordial power spectrum, consistent with the current constraints from Planck~\cite{Planck:2015}. To describe the redshift dependence of the pSZ effect, we introduce a window function $W(\chi_e, \bar{\chi}_e)$ that gives the electron density in a set of redshift bins centered on $\chi_e = \bar{\chi}_e$
\begin{equation}
\delta({\bf \hat{n}}_e, \bar{\chi}_e)=\int d\chi_e W(\chi_e, \bar{\chi}_e)\delta({\bf \hat{n}}_e, \chi_e).
\end{equation}
In this work, we use a top-hat window function normalized to unity, $\int_0^{\chi_\infty} d\chi W(\chi, \bar{\chi})=1$, and we consider six redshift bins of equal width, covering the range $0 < z < 6$. The redshift coverage for each bin configuration is shown in table~\ref{tab:bins}. 

\begin{table}[]
\centering
\begin{tabular}{ccc}
\hline \hline
\multicolumn{1}{c}{$\bar{\chi}_e$} & \multicolumn{1}{c}{$\chi_e$} & \multicolumn{1}{c}{$z_e$} \\ \hline
$\bar{\chi}_e=0.16$ & $0.00<\chi_e<0.32$ & $0.00<z_e<0.35$ \\ 
$\bar{\chi}_e=0.48$ &$0.32<\chi_e<0.64$ & $0.35<z_e<0.78$ \\ 
$\bar{\chi}_e=0.80$ &$0.64<\chi_e<0.96$ & $0.78<z_e<1.37$ \\ 
$\bar{\chi}_e=1.12$ &$0.96<\chi_e<1.28$ & $1.37<z_e<2.22$ \\ 
$\bar{\chi}_e=1.44$ &$1.28<\chi_e<1.60$ & $2.22<z_e<3.59$ \\ 
$\bar{\chi}_e=1.76$ &$1.60<\chi_e<1.92$ & $3.59<z_e<6.00$ \\ 
\hline \hline
\end{tabular}
\caption{Redshift bins: six equally spaced redshift bins from here ($z=0$) to reionization ($z=6$). The comoving distance is given in units of $H_0^{-1}$.}
\label{tab:bins}
\end{table}

We first consider the cross-correlation of the Stokes parameters $Q \pm i U$ with the density field $\delta$ which traces the electron distribution, given by 
\begin{equation}
\begin{split}
	\bang{ (Q \pm i U)^{\text{pSZ}} ({\bf \hat{n}}_e) \delta({\bf \hat{n}}'_e, & \bar{\chi}_e)} = - \frac{\sqrt{6} \sigma_T}{10}  \int d\chi_e \, a_e \, \bar{n}_e (\chi_e)  \sum_{m=-2}^{2}   \left._{\pm 2}Y_{2 m}\right. ({\bf \hat{n}}_e)   \\
	& \times \int d\chi'_e W(\chi'_e, \bar{\chi}_e) \bang{ (1+\delta ({\bf \hat{n}}_e,\chi_e)) \ q^{m}_{\rm eff} ({\bf \hat{n}}_e, \chi_e)  \delta({\bf \hat{n}}_e', \chi_e')}.
\end{split}
\label{eq:QiUdeltacorrelation}
\end{equation}
A crucial point for our analysis is that we can isolate large scale inhomogeneities by treating $q^{m}_{\rm eff}$ as a slowly varying deterministic field, and treating $\delta$ as a stochastic field with variations on small scales. We'll now formalize this split between large and small scale contributions to see how it helps achieve our goal of unlocking large scale information. We begin by defining a long and short wavelength decomposition of $\Psi$, 
\begin{equation}\label{eq:longshortpsi}
\Psi_i^\mathcal{L} ({\bf x}) = \int \frac{d^3 k}{(2 \pi)^3} \mathcal{L} (k) \Psi_i ({ \bf k}) e^{i {\bf k} \cdot {\bf x}}, \ \ \ \   \Psi_i^\mathcal{S} ({\bf x}) = \int \frac{d^3 k}{(2 \pi)^3} \mathcal{S} (k) \Psi_i ({ \bf k}) e^{i {\bf k} \cdot {\bf x}} ,
\end{equation}
where $\mathcal{L} (k) + \mathcal{S} (k) = 1$. For instance, we may choose $\mathcal{L} (k) = e^{- k^2 / 2 k_{*}^2 }, \ \mathcal{S} (k) = 1 - e^{- k^2 / 2 k_{*}^2 }$. We will generally assume $k_*\sim (100\ \text{Mpc})^{-1}$, but the results do not depend on the exact value of $k_*$. The signal described in this section is sensitive only to the deterministic long field formed by scales larger than $100\ \text{Mpc}$  (as illustrated later in figure~\ref{fig:clq}) while the noise described in section~\ref{sec:noise} depends mainly on the stochastic short field formed by scales smaller than $100\ \text{Mpc}$.

The decomposition in~\eqref{eq:longshortpsi} implies a similar long-short split for the quadrupole and density fields, valid in the linear regime
\begin{align}
{q^{m}_{\rm eff}}({\bf \hat{n}}_e,\chi_e) =  & \ {q^{m}_{\rm eff}}^\mathcal{L} ({\bf \hat{n}}_e,\chi_e) +  {q^{m}_{\rm eff}}^\mathcal{S} ({\bf \hat{n}}_e,\chi_e), \no \\
 \delta({\bf \hat{n}}_e,\chi_e) = & \ \delta^\mathcal{L} ({\bf \hat{n}}_e,\chi_e) + \delta^\mathcal{S} ({\bf \hat{n}}_e,\chi_e).
\end{align}
Substituting this expansion into the cross-correlation in the second line of~\eqref{eq:QiUdeltacorrelation} we obtain
\begin{align}
	\ang{(1+\delta)\ q_\text{eff}\ \delta'} & = \ \ang{(1 + \delta^\mathcal{L} + \delta^\mathcal{S})\ (q_\text{eff}^\mathcal{L} + q_\text{eff}^\mathcal{S}) \ ({\delta'}^\mathcal{L} +{\delta'}^\mathcal{S})} \no \\
\begin{split}
	& = \ q_\text{eff}^\mathcal{L}{\delta'}^\mathcal{L}
	+ q_\text{eff}^\mathcal{L}\delta^\mathcal{L}{\delta'}^\mathcal{L} 
	+ \ang{ q_\text{eff}^\mathcal{S} {\delta'}^\mathcal{S}} \\
      & \quad + q_\text{eff}^\mathcal{L} \ang{\delta^\mathcal{S}{\delta'}^\mathcal{S}}
	+ \delta^\mathcal{L}\ang{ q_\text{eff}^\mathcal{S} {\delta'}^\mathcal{S}}
	+ {\delta'}^\mathcal{L}\ang{ q_\text{eff}^\mathcal{S} {\delta}^\mathcal{S}}. \\
	\end{split}
	\label{eq:qdelta_longshort}
\end{align} 
The main point of this analysis is that if we want to learn information about large scale inhomogeneities, the ensemble average in eq.~\eqref{eq:QiUdeltacorrelation} should only be taken over small scales, leaving large scales as a fixed deterministic field. Above, we have used our assumption of Gaussian fields with zero mean to set to zero any one-point and three-point correlation functions for the short modes. Of the remaining terms, those involving only long modes are a deterministic contribution that shows up only on large angular scales. These terms will be negligible compared to the primary CMB, and are neglected below. The statistically isotropic cross-power between short modes of the density and quadrupole field do not contribute to the signal of interest. Of the three statistically anisotropic terms, only the term $q_\text{eff}^\mathcal{L} \ang{\delta^\mathcal{S}{\delta'}^\mathcal{S}}$ in~\eqref{eq:qdelta_longshort} is significant since the quadrupole field is primarily made up of long-wavelength modes.~\footnote{We can estimate the effect of adding bias in as follows. In a purely local bias model to leading order we can assume that $\delta_e = b_1 \delta + b_2 \delta^2 + \ldots$. With this assumption, the statistically anisotropic component is given by
\begin{eqnarray*}
\ang{(1+\delta_e)\ q_\text{eff}\ \delta'} &\subset& q_\text{eff}^\mathcal{L} \left[ 
b_1^2 \ang{{\delta}^\mathcal{S} {\delta'}^\mathcal{S}} 
+ b_2 \ang{{\delta'}^\mathcal{S} {\delta'}^\mathcal{S}} 
+ b_2^2  \ang{{\delta'}^\mathcal{S} {\delta'}^\mathcal{S} {\delta}^\mathcal{S} {\delta}^\mathcal{S}} \right. \\
&+& \left. b_1 b_2 {\delta'}^\mathcal{L} \ang{{\delta}^\mathcal{S} {\delta}^\mathcal{S}} 
+ b_1 b_2 {\delta}^\mathcal{L} \ang{{\delta'}^\mathcal{S} {\delta'}^\mathcal{S}}
+2 b_1 b_2 {\delta}^\mathcal{L} \ang{{\delta}^\mathcal{S} {\delta'}^\mathcal{S}}
+\mathcal{O}({{\delta}^\mathcal{L}}^2)
\right]
\end{eqnarray*}
On linear scales, the density contrast is small (${\delta}^\mathcal{L} \ll 1$) implying that we can safely neglect the terms in the second line of this equation. What remains is again small scale power modulated by $q_\text{eff}$. Roughly speaking, as long as $b_1^2 (1 + b_2/b_1^2) > 1$ there will be an enhancement in small-scale power over what we have assumed in the main text. Generally speaking, $b_1 \geq 1$, and on non-linear scales one expects $b_2< 0$ from the injection of energy due to baryonic feedback effects. At the resolutions assumed in our signal to noise estimates in section~\ref{sec:StoN} ($\ell_{\rm max}=3000$), we do not expect to probe the highly non-linear regime in all but the first redshift bin (for example, scales of 1 Mpc subtend an angle less than $\pi / \ell_{\rm max}$ for $z > 0.3$). Therefore, we expect that incorporating the linear bias term is sufficient, and we defer a more careful treatment of nonlinear bias (and more realistic tracers!) to future work.} \emph{This power asymmetry is our desired signal.} Using this, we approximate the correlation in eq.~\eqref{eq:QiUdeltacorrelation} as 
\begin{equation}
\begin{split}
	\bang{ (Q \pm i U)^{\text{pSZ}} ({\bf \hat{n}}_e) \delta({\bf \hat{n}}'_e, \bar{\chi}_e)} & = - \frac{\sqrt{6} \sigma_T}{10} \int d\chi_e  a_e \bar{n}_e (\chi_e)  \sum_{m=-2}^{2} \ q^{m}_{\rm eff} ({\bf \hat{n}}_e, \chi_e)  \left._{\pm 2}Y_{2 m}\right. ({\bf \hat{n}}_e)  \\
	& \times \int d\chi'_e\ W(\chi'_e, \bar{\chi}_e) \ \bang{ \delta ({\bf \hat{n}}_e,\chi_e) \   \delta({\bf \hat{n}}_e', \chi_e')} + \text{isotropic},
\end{split}
\label{eq:def-cross-correlation-pol-delta}
\end{equation}
where we've suppressed the $\mathcal{S}$ and $\mathcal{L}$ superscripts, and the density autocorrelation is given in equations~\eqref{eq:angdelta},\eqref{eq:Cldeltadelta_chi}. This results motivates an estimator of the remote quadrupole field by cross correlating $(Q \pm i U)$ and $\delta$. However in practice one obtains a better estimator by first splitting the polarisation field into E and B modes, which we will do in the next section.
%We characterize the power asymmetry by computing the \emph{power multipoles} obtained by projecting eq.~\eqref{eq:QiUdeltacorrelation} onto spin-weighted spherical harmonics $_{\pm 2}Y_{\ell m}({\bf \hat{n}}_e) $. Under the assumption of random Gaussian fields, the power multipoles are in one-to-one correspondence to the moments of the quadrupole field. 

\subsection{Signal calculation}

The polarisation $(Q \pm i U)^{\text{pSZ}} ({\bf \hat{n}}_e)$ contains both $E$-mode and a $B$-mode components (even from pure scalar perturbations). However, the $E$ and $B$ mode background, from which we wish to distinguish our signal, have drastically different magnitudes over a large range of angular scales; see Figure~\ref{fig:ClEEpSZ}. To maximize the signal to noise, it is therefore useful to define a set of estimators that are based on the information in the $E$ and $B$ mode polarization separately. To do so, we define a set of two new spin-2 fields ${}_{\pm2}X^{\text{pSZ}}$ based on the scalar $X=E,B$ fields:
\be
 {}_{\pm2}X^{\text{pSZ}}({\bf \hat{n}}_e) = \sum_{\ell m} X^{\text{pSZ}}_{\ell m} \  {}_{\pm2}Y_{LM}^{*}({\bf \hat{n}}_e)
\ee
In each redshift bin we calculate the expected correlation
\begin{equation} \label{eq:powermultipoles}
	{}_{\pm2}a^X_{LM} \equiv \int d^2 \hat{n}_e\ {}_{\pm2}Y_{LM}^{*}({\bf \hat{n}}_e) \bang{  {}_{\pm2}X^{\text{pSZ}}({\bf \hat{n}}_e) \delta({\bf \hat{n}}_e, \bar{\chi}_e)} . 
\end{equation}
Importantly, the above quantity allows us to isolate the statistically anisotropic term in eq.~\eqref{eq:def-cross-correlation-pol-delta}, which makes it possible to measure the effective quadrupole. Since the effective quadrupole is related to the primordial potential as in eq.~\eqref{eq:initial-density-to-effective-quadrupole}, this provides a  way to measure large scale inhomogeneities. 

For the correlator of ${}_{\pm2}X^{\text{pSZ}}$ with matter, assuming only scalar perturbations we obtain, using Eq.~\eqref{eq:pSZB1} and~\eqref{eq:pSZE1} 
\begin{align}\label{eq:corrXdelta}
\bang{  {}_{\pm2}X^{\text{pSZ}}({\bf \hat{n}}_e) \delta({\bf \hat{n}}_e, \bar{\chi}_e)} = & \frac{\sqrt{6} \ \sigma_T}{10}  \int d\chi_e \  a_e \bar{n}_e (\chi_e) \sum_{\ell_1,m_1}\sum_{\ell_2,m_2}\sum_{\ell,m}  {}_{\pm2}Y_{\ell_1m_1}({\bf \hat{n}}_e) Y_{\ell_2 m_2}({\bf \hat{n}}_e) \no \\
& \frac{1}{2i} \left( 1 \pm (-1)^{\ell_1+\ell_2+\ell}\right) \sqrt{\frac{(2 \ell_1+1) (2\ell_2+1) (2\ell+1)}{4 \pi}} (-1)^{m_1} \no \\
& \times \left(\begin{array}{ccc} \ell_1 & \ell_2 & \ell \\ -m_1 & -m_2 & m \end{array} \right) \left(\begin{array}{ccc} \ell_1 & \ell_2 & \ell \\ -2 & 0 & 2 \end{array} \right) C_{\ell_2}^{\delta\delta} (\chi_e) a_{\ell m}^{q} (\chi_e).
\end{align}
%Here the sign in ${}_{\pm2}B^{\text{pSZ}}$ indicates the spin sign of the spin 2 remote quadrupole component $a_{\ell m}^{q\pm}$. 
%In this paper we are focussing on the scalar contribution to the quadrupole for which $a_{\ell m}^{q+} = a_{\ell m}^{q-}$ and the $b^X_{LM}$ are zero. 
where in the $(1 \pm (-1)^{\ell_1+\ell_2+\ell})$ factor the positive sign is for $X=E$ and the negative sign for $X=B$. Plugging the correlator in Eq.~\eqref{eq:powermultipoles} we obtain
\begin{align} \label{eq:powermultipoles2}
	{}_{\pm2}a^X_{LM}(\bar{\chi}_e) = &  \frac{\sqrt{6} \ \sigma_T}{10}  \int d\chi_e \  a_e \bar{n}_e (\chi_e) \int d\chi'_e\ W(\chi'_e, \bar{\chi}_e)  \no \\
& \sum_{\ell_1\ell_2} \frac{1}{2i} \left( 1 \pm (-1)^{\ell_1+\ell_2+L}\right) \frac{(2 \ell_1+1) (2\ell_2+1) }{4 \pi} \no \\
& \times \left(\begin{array}{ccc} \ell_1 & \ell_2 & L \\ -2 &  0 & 2 \end{array} \right) \left(\begin{array}{ccc} \ell_1 & \ell_2 & L \\\pm 2 & 0 & \mp 2 \end{array} \right) C_{\ell_2}^{\delta\delta} (\chi_e,\chi'_e) a_{LM}^q (\chi_e)
\end{align}
% The equations for $e^E_{LM}$ are almost identical and we get
% \begin{align} \label{eq:powermultipoles3}
% 	e^E_{LM}(\bar{\chi}_e) = &  \frac{\sqrt{6} \ \sigma_T}{10}  \int d\chi_e \  a_e \bar{n}_e (\chi_e) \int d\chi'_e\ W(\chi'_e, \bar{\chi}_e)  \no \\
% & \sum_{\ell_1\ell_2} \frac{1}{2} \left( 1 + (-1)^{\ell_1+\ell_2+\ell}\right) \frac{(2 \ell_1+1) (2\ell_2+1) }{4 \pi} \no \\
% & \times \left(\begin{array}{ccc} \ell_1 & \ell_2 & L \\ -2 &  0 & 2 \end{array} \right) \left(\begin{array}{ccc} \ell_1 & \ell_2 & L \\ 0 & 0 & 0 \end{array} \right) C_{\ell_2}^{\delta\delta} (\chi_e,\chi'_e) a_{LM}^q (\chi_e)
% \end{align}

Finally, we define a set of scalar multipole moments
\begin{eqnarray}
e_{LM} &=& \frac{1}{2} \left( {}_{+ 2} a^E_{LM} + {}_{- 2} a^E_{LM} \right) ,\\
b_{LM} &=& \frac{1}{2i} \left( {}_{+ 2} a^B_{LM} - {}_{- 2} a^B_{LM} \right),
\end{eqnarray}
Inserting $C_{L'}^{\delta\delta}(\chi_e,\chi'_e)$ from Eq.~\ref{eq:Cldeltadelta_chi} and applying the Limber approximation we obtain
\begin{align} \label{eq:powermultipoles3}
	b_{LM}(\bar{\chi}_e) = &  \frac{\sqrt{6} \ \sigma_T}{10}  \sum_{\ell_1\ell_2} \frac{1}{4} \left( 1 - (-1)^{\ell_1+\ell_2+L}\right)^2 \frac{(2 \ell_1+1) (2\ell_2+1) }{4 \pi} \left(\begin{array}{ccc} \ell_1 & \ell_2 & L \\ -2 &  0 & 2 \end{array} \right)^2 \no \\
& \int \frac{d\chi_e}{\chi_e^2} \  a_e \bar{n}_e (\chi_e) W(\chi_e, \bar{\chi}_e)  a_{LM}^q (\chi_e) P_\delta \left(\frac{\ell_2+1/2}{\chi_e},\chi_e \right) \\
%& = \ 0
\end{align}
and
\begin{align} \label{eq:powermultipoles4}
	e_{LM}(\bar{\chi}_e) = &  \frac{\sqrt{6} \ \sigma_T}{10}  \sum_{\ell_1\ell_2} \frac{1}{4} \left( 1 + (-1)^{\ell_1+\ell_2+L}\right)^2 \frac{(2 \ell_1+1) (2\ell_2+1) }{4 \pi} \left(\begin{array}{ccc} \ell_1 & \ell_2 & L \\ -2 &  0 & 2 \end{array} \right)^2 \no \\
& \int \frac{d\chi_e}{\chi_e^2} \  a_e \bar{n}_e (\chi_e) W(\chi_e, \bar{\chi}_e)  a_{LM}^q (\chi_e) P_\delta \left(\frac{\ell_2+1/2}{\chi_e},\chi_e \right)
\end{align}
%Similar expressions apply for tensor sourced B-mode remote quadrupole fields $b^X_{LM}$.
We thus find that the large-scale quadrupole induces E-mode and B-mode modulations of the correlation of CMB and matter, both of which are proportional to the multipoles $a_{LM}^q$. We can thus use this correlation as a probe of $a_{LM}^q$.

\subsection{Variance calculation} \label{sec:noise}

In this section we calculate the Gaussian variance of the estimator $\hat{e}_{LM}$ and $\hat{e}_{LM}$ in the absence of the pSZ signal induced by the remote quadrupole field. Our estimator (indicated notationally by the overhead) is defined as
\begin{eqnarray}
\hat{e}_{LM} &=& \frac{1}{2} \left( {}_{+ 2} \hat{a}^E_{LM} + {}_{- 2} \hat{a}^E_{LM} \right) ,\\
\hat{b}_{LM} &=& \frac{1}{2i} \left( {}_{+ 2} \hat{a}^B_{LM} - {}_{- 2} \hat{a}^B_{LM} \right),
\end{eqnarray}
with
\begin{equation}% \label{eq:powermultipoles}
	{}_{\pm2}\hat{a}^X_{LM} \equiv \int d^2 \hat{n}_e\ {}_{\pm2}Y_{LM}^{*}({\bf \hat{n}}_e) {}_{\pm2}X^{\text{pSZ}}({\bf \hat{n}}_e) \delta({\bf \hat{n}}_e, \bar{\chi}_e) . 
\end{equation}
To make the separation between large scales and small scales concrete, we filter both the CMB map $X$ and the matter map $\delta$ with a high pass filter $\ell > \ell_{\rm min}$. 

We now calculate the variance of this estimator, which corresponds to an accidental power asymmetry in the cross-correlation in the absence of the remote quadrupole field. 
Starting with the $b_{LM}$ , we compute:
 \begin{equation}
 \begin{split}
	\bang{\hat{b}^*_{LM} (\bar{\chi}_e) \hat{b}_{LM} (\bar{\chi}_e)} & = \frac{1}{4} \bang{ ({}_{+ 2} \hat{a}^{B*}_{LM} - {}_{- 2} \hat{a}^{B*}_{LM})  ({}_{+ 2} \hat{a}^B_{LM} - {}_{- 2} \hat{a}^B_{LM} )}
\end{split}
\label{eq:noisedef}
\end{equation}
where $\bar{\chi}_e$ represents the center of the redshift bin. This variance quantifies the chance power asymmetry that is present in the statistically isotropic contribution to $E,B$, which is sensitive mainly to small scales. For example the first term gives
 \begin{equation}
 \begin{split}
\bang{ {}_{+ 2} \hat{a}^{B*}_{LM}  {}_{+ 2} \hat{a}^B_{LM} } &=  \int d^2{\bf \hat{n}}_e d^2{\bf \hat{n}}'_e\ {}_{+ 2} Y_{LM}({\bf \hat{n}}_e) \ {}_{+ 2} Y^*_{LM}({\bf \hat{n}}'_e) \bang{ {}_{+ 2} B({\bf \hat{n}}_e)\ \delta({\bf \hat{n}}_e, \bar{\chi}_e)\ {}_{+ 2} B({\bf \hat{n}}'_e)\ \delta({\bf \hat{n}}'_e, \bar{\chi}_e)}\\
&\simeq  \int d^2{\bf \hat{n}}_e d^2{\bf \hat{n}}'_e\ {}_{+ 2} Y_{LM}({\bf \hat{n}}_e) \ {}_{+ 2} Y^*_{LM}({\bf \hat{n}}'_e) \bang{ {}_{+ 2} B({\bf \hat{n}}_e) \ {}_{+ 2}  B({\bf \hat{n}}'_e)} \bang{\delta({\bf \hat{n}}_e, \bar{\chi}_e)\  \delta({\bf \hat{n}}'_e, \bar{\chi}_e)},
\end{split}
\end{equation}
where in the second line we have dropped the two largely subdominant cross correlation terms.

The binned matter density power spectrum is given by 
\begin{equation}
	\bang{ \delta({\bf \hat{n}}_e, \bar{\chi}_e)\ \delta({\bf \hat{n}}'_e, \bar{\chi}_e)} = \sum_{\ell,m} C_\ell^{\delta \delta} \ Y^*_{\ell m}({\bf \hat{n}}_e)  \ Y_{\ell m}({\bf \hat{n}}'_e), \label{eq:Cldd1}
\end{equation}	
with
\begin{align}
	C_\ell^{\delta\delta}(\bar{\chi}_e)  = & \int d\chi_e \ W(\chi_e, \bar{\chi}_e) \int d\chi'_e \ W(\chi'_e, \bar{\chi}_e) \ C_\ell^{\delta\delta}(\chi_e,\chi'_e) \no \\
	= & \int dk \frac{2k^2}{\pi} \int d\chi_e \sqrt{P_\delta(k,\chi_e)}W(\chi_e, \bar{\chi}_e)j_\ell(k\chi_e) \int d\chi'_e \sqrt{P_\delta(k,\chi'_e)}W(\chi'_e, \bar{\chi}_e)j_\ell(k\chi'_e) \no \\
	\simeq & \int\frac{dk}{\ell+1/2} W^2\left(\frac{\ell+1/2}{k}, \bar{\chi}_e\right)P_\delta\left(k,\frac{\ell+1/2}{k}\right),  \label{eq:Cldd2}
\end{align}
where we used the expression for $C_\ell^{\delta\delta}(\chi_e,\chi'_e)$ from~\eqref{eq:Cldeltadelta_chi}, and the Limber approximation in the last line.

The CMB power spectra $C_\ell^{XX}$ are given by
\begin{equation}
 \begin{split}
\bang{{}_{\pm2}  X({\bf \hat{n}}_e) {}_{\pm2} X({\bf \hat{n}}'_e)} = & \sum_{\ell,m} C_\ell^{XX} \ {}_{\pm 2}Y^*_{\ell m}({\bf \hat{n}}_e)  \ {}_{\pm 2}Y_{\ell m}({\bf \hat{n}}'_e), 
\end{split}
\end{equation}
The contributions to the $E$-mode and $B$-mode power spectra at the high $\ell$ of our interest are
\begin{eqnarray} \label{eq:ClEEtotal}
	C_\ell^{EE} &=& C_\ell^{EE,\text{lensed}} +  C_\ell^{EE,\text{pSZ}}, \\
	C_\ell^{BB} &=& C_\ell^{BB,\text{lensed}} +  C_\ell^{BB,\text{pSZ}},
\end{eqnarray}
for which we can recall the expressions for $C_\ell^{EE,\text{pSZ}}$, $C_\ell^{EE,\text{lensed}}$, $C_\ell^{BB,\text{lensed}}$, and $C_\ell^{BB,\text{pSZ}}$ from equations~\eqref{eq:ClEpSZapp}, \eqref{eq:ClBpSZ1}, and \eqref{eq:lensing_high_ell_approx}. The various terms are plotted in Figure~\ref{fig:ClEEpSZ}.

Plugging these expressions into the first term of the variance we obtain 
 \begin{equation}
 \begin{split}
\bang{ {}_{+ 2} \hat{a}^{B*}_{LM}  {}_{+ 2} \hat{a}^B_{LM} } &= \sum_{\ell,\ell',m,m'} C_\ell^{BB}\ C_{\ell'}^{\delta\delta}(\bar{\chi}_e) \frac{(2L+1) (2\ell+1)(2\ell'+1)}{4\pi} \\
&\times \left(\begin{array}{ccc} L & \ell & \ell' \\  M & -m & -m' \end{array} \right)^2 \left(\begin{array}{ccc} L & \ell & \ell' \\  +2 & -2 & 0 \end{array} \right)^2
\end{split}
\end{equation}

Including all permuations and using the orthogonality relation of the 3j-symbols the final result for the B-mode variance is
  \begin{equation}
 \begin{split}
	\bang{\hat{b}_L(\bar{\chi_e})^2} = \frac{1}{2} \sum_{\ell,\ell'=\ell_\text{min}}^{\ell_\text{max}} C_\ell^{BB}\ C_{\ell'}^{\delta\delta}(\bar{\chi}_e)  \frac{(2\ell+1)(2\ell'+1)}{4\pi}\left(\begin{array}{ccc} L & \ell & \ell' \\  +2 & -2 & 0 \end{array} \right)^2 %\left(1 + (-1)^{L+\ell+\ell'}\right)
	\label{eq:noisesumB}
\end{split}
\end{equation}
and for the E-modes
  \begin{equation}
 \begin{split}
	\bang{\hat{e}_L(\bar{\chi_e})^2} = \frac{1}{2} \sum_{\ell,\ell'=\ell_\text{min}}^{\ell_\text{max}} C_\ell^{EE}\ C_{\ell'}^{\delta\delta}(\bar{\chi}_e)  \frac{(2\ell+1)(2\ell'+1)}{4\pi}\left(\begin{array}{ccc} L & \ell & \ell' \\  +2 & -2 & 0 \end{array} \right)^2 %\left(1 + (-1)^{L+\ell+\ell'}\right).
	\label{eq:noisesumE}
\end{split}
\end{equation}
Note that the 3-$j$ symbols are only nonzero for $|\ell-L| \leq \ell' \leq \ell+L$. Just as for the signal, the lower and upper bounds on the sum have been introduced to represent the experimental filtering and resolution scales.

\section{Experimental forecast}
\label{sec:detection}

In this section, we explore the sensitivity and resolution requirements to detect the pSZ signal, and compare with what is attainable in future experiments. Our strategy will be to assume a cosmic-variance limited measurement out to a fiducial choice for $\ell_{\rm max}=3000$, determine the required resolution and sensitivity of a CMB polarization experiment and galaxy survey to achieve this, and then compute the signal-to-noise ratio (SNR) in the 6-redshift bin configuration. %Here, we consider galaxy shot noise, CMB instrumental noise and angular resolution. A more realistic treatment including systematics and foregrounds is necessary for a complete assessment of how well we might measure the pSZ signal. As this is most effectively done in the context of a specific instrument, we do not venture to do so here.

\subsection{Experimental requirements}

Assuming that the instrumental noise is a uniform Gaussian random field, the total observed CMB $E$-mode and $B$-mode power spectra can be written as a sum of three different contributions:
\begin{align}
	C^{EE}_{\ell} =& \left( C_{\ell}^{EE,\text{lensed}} + C_{\ell}^{EE,\text{pSZ}} + N^{EE} \right) \exp\left[ \frac{\ell (\ell+1) \theta^2_{\text{FWHM}}}{8 \ln 2} \right], \\
	C^{BB}_{\ell} =& \left( C_{\ell}^{BB,\text{lensed}} + C_{\ell}^{BB,\text{pSZ}} + N^{BB} \right) \exp\left[ \frac{\ell (\ell+1) \theta^2_{\text{FWHM}}}{8 \ln 2} \right],
\end{align}
where $\theta_{\text{FWHM}}$ is the full width at half maximum (expressed in radians), and $C_{\ell}^{EE,\text{lensed}}$, $C_{\ell}^{EE,\text{pSZ}}$ and $N^{EE}$ denote the lensed primary $E$-mode power spectrum, the high-$\ell$ pSZ power, and the (gaussian, white) instrumental noise respectively; the corresponding quantities are also defined for the $B$-modes.

For a given $\ell_{\text{max}}$, the corresponding angular resolution (in radians) required is
\begin{equation}
	\theta_{\text{FWHM}} = \sqrt{ \frac{8 \ln 2}{\ell_{\text{max}}(\ell_{\text{max}}+1)}}.
\end{equation}
For $\ell_{\text{max}}$ this translates to $\theta_{\text{FWHM}} = 2.7 \ {\rm arcmin}$. The required sensitivity can then be set by matching the power of the noise $N^{XX}$ to the power of the other contributions to the $X=E,B$ modes at $\ell_{\text{max}}$:
\begin{eqnarray}
	N^{XX} = C_{\ell_{\text{max}}}^{XX,\text{lensed}} + C_{\ell_{\text{max}}}^{XX,\text{pSZ}}.
\end{eqnarray}
For $\ell_{\text{max}}=3000$, and using the $E$ and $B$ mode power spectra shown in Figure~\ref{fig:ClEEpSZ}, this translates to $N^{EE} \simeq 3 \ \mu{\rm K} \ {\rm arcmin}$ and $N^{BB} \simeq 0.3 \ \mu{\rm K} \ {\rm arcmin}$. A Stage 4 CMB experiment (CMB S4) is aiming for an angular resolution between 1 and 3 arcmin with a noise level of 1 $\mu$K arcmin~\cite{CMBS42016} in temperature, and a factor of $\sqrt{2}$ higher in polarization. Therefore, CMB S4 would likely have adequate angular resolution and adequate noise for the $E$-mode signal, but would fall short by a factor of $\sim 4$ to reach the cosmic variance limit for the $B$-mode signal at $\ell_{\text{max}}=3000$. 
%
% (corresponding to 
%
%
%%$\sigma \sim .4 - .6 \ \mu {\rm K} \ \theta_{\rm FWHM}^{-1}$
%$\sigma \sim .5 - .7 \ \mu {\rm K} \ \theta_{\rm FWHM}^{-1}$ 
%in the two resolution scenarios of table~\ref{Table:sensitivity-reqs}). CMB S4 would therefore provide adequate resolution for either case, but the sensitivity is just on the verge of being sufficient for the low-resolution scenario, and falls short of the high-resolution scenario by roughly an order of magnitude. 
%%but the sensitivity would only be sufficient for the low-resolution scenario, falling short of the high-resolution scenario by a factor of roughly seven. 
%
%\begin{table}
%	\centering
%	\begin{tabular}{cccc}
%	\hline \hline 
%	$\ell_{\text{max}}$ & $\theta_{\text{FWHM}}$ (arcmin) & $\sigma_{N/F}$ ($\mu$K $\theta_{\rm FWHM}^{-1}$) & $N_g (\bar{\chi}_e)$ (arcmin$^{-2}$) \\
%	\hline 
%	%3000 & 2.70 & 1.06 & 14, 26, 82, 255, 697, 1662 \\
%	4000 & 2.02 & 0.34 & 27, 42, 116, 377, 1174, 3042 \\
%	6000 & 1.35 & 0.06 & 69, 87, 201, 630, 2298, 7169 \\
%	\hline \hline
%	\end{tabular}
%	\caption{Sensitivity requirements of an experiment for two different scenarios with six redshift bins.}
%	\label{Table:sensitivity-reqs}
%\end{table}

Our analysis also involves the density of free electrons $\delta_e$. We will assume that $\delta_e$ is traced by the galaxy number density $\delta_g$. A more careful analysis should include the bias between the free electrons and galaxy number distributions; see Sec.~\ref{eq:longshortsplitsec} for a discussion. Galaxy number density is affected by shot noise due to discrete sampling of galaxies, yielding a measured angular power spectrum of:
\begin{equation}
	C_\ell^{\delta\delta}(\bar{\chi}_e) \rightarrow C_\ell^{\delta\delta}(\bar{\chi}_e) + \frac{1}{\Delta N(\bar{\chi}_e)},
\end{equation}
where $\Delta N(\bar{\chi}_e)$ is the  number of galaxies per steradian in the redshift bin centered on $\bar{\chi}_e$.
The galaxy densities necessary for $C_\ell^{\delta\delta}(\bar{\chi}_e)$ to dominate the shot noise in each redshift bin at $\ell_{\text{max}}=3000$ is $\Delta N(\bar{\chi}_e) = \{14, 26, 82, 255,697,1661\} \ {\rm arcmin}^{-2}$. This can be compared to the capabilities of Euclid~\cite{Laureijs2011} and LSST~\cite{LSSTScienceCollaboration2009}, which expect to reach a total galaxy number density of $N_g^{\text{Euclid}} \sim 30 \ {\rm arcmin}^{-2}$ and $N_g^{\text{LSST}} \sim 130 \ {\rm arcmin}^{-2}$ respectively. Neglecting the distribution over redshift, this would be enough to cover the two or three redshift bins closest to us.

\subsection{Signal-to-Noise ratio} \label{sec:StoN}

Having calculated the expected signal $b_{L M}$ and  $e_{L M}$, as well as the estimator variance we obtain the signal-to-noise ratio per mode by
\be
\label{eq:qSN}
\frac{S}{N}(L,\bar{\chi}_e) = \left[ \frac{f_{\rm sky}}{2} \left( \frac{C_{L}^{X}(\bar{\chi}_e)}{N_{L}^{X}(\bar{\chi}_e)} \right)^2 \right]^{1/2}.
\ee
Here $C_{L}^{X}(\bar{\chi_e}) = \bang{x_{L M}^2(\bar{\chi_e})}$ is the signal power spectrum for $x=e,b$, where $x_{L M}$ is given in Eq.~\eqref{eq:powermultipoles3} and Eq.~\eqref{eq:powermultipoles4}. The noise $N_{L}^{X}(\bar{\chi_e}) = \bang{\hat{x}_L(\bar{\chi_e})^2}$ was calculated in Eq.~\eqref{eq:noisesumB} and Eq.~\eqref{eq:noisesumE}.

\begin{figure}[htbp]
	\begin{center}
		\subfigure{\includegraphics[width=7.5cm]{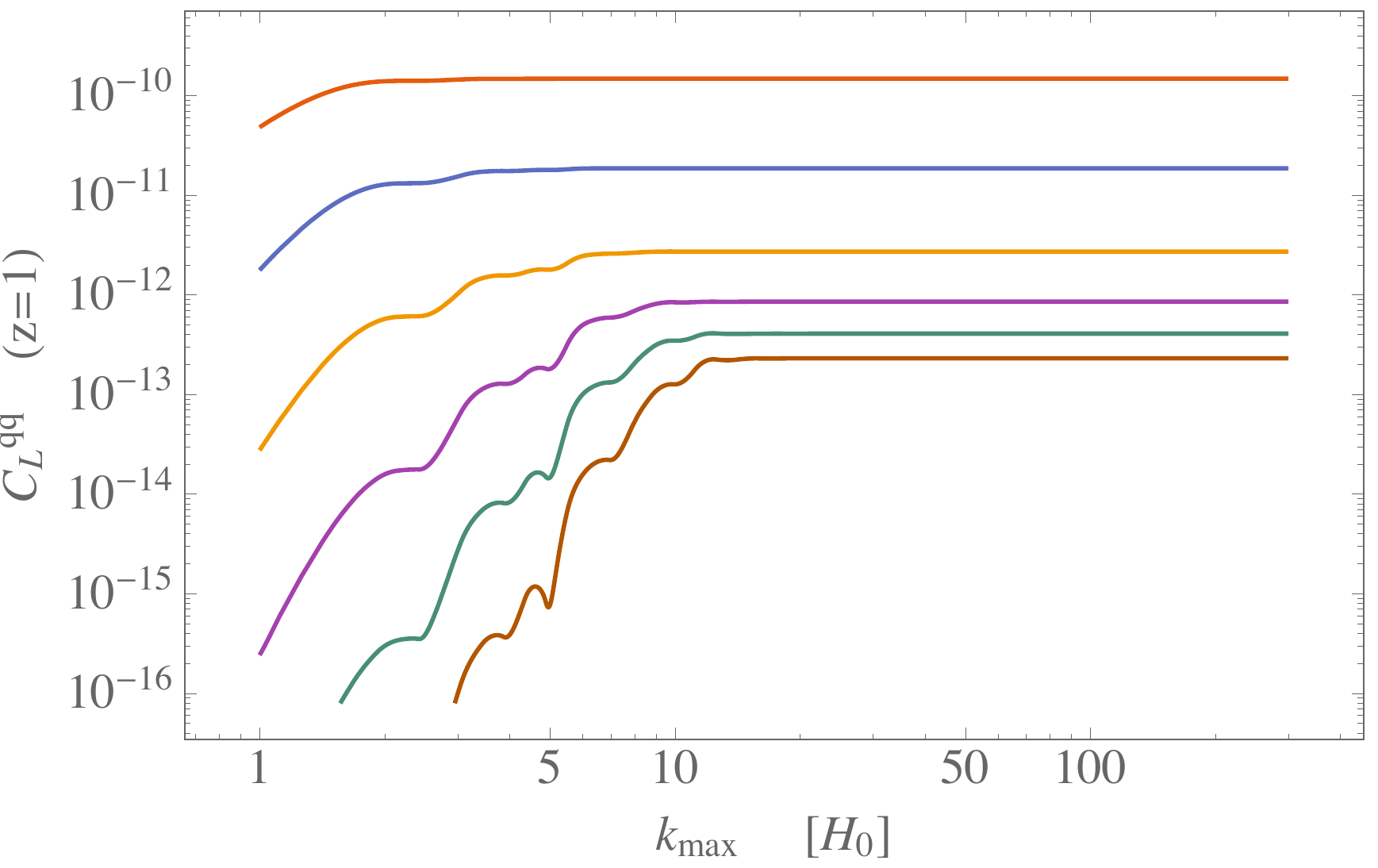}}
		\hspace{.2cm}
		\subfigure{\includegraphics[width=7.5cm]{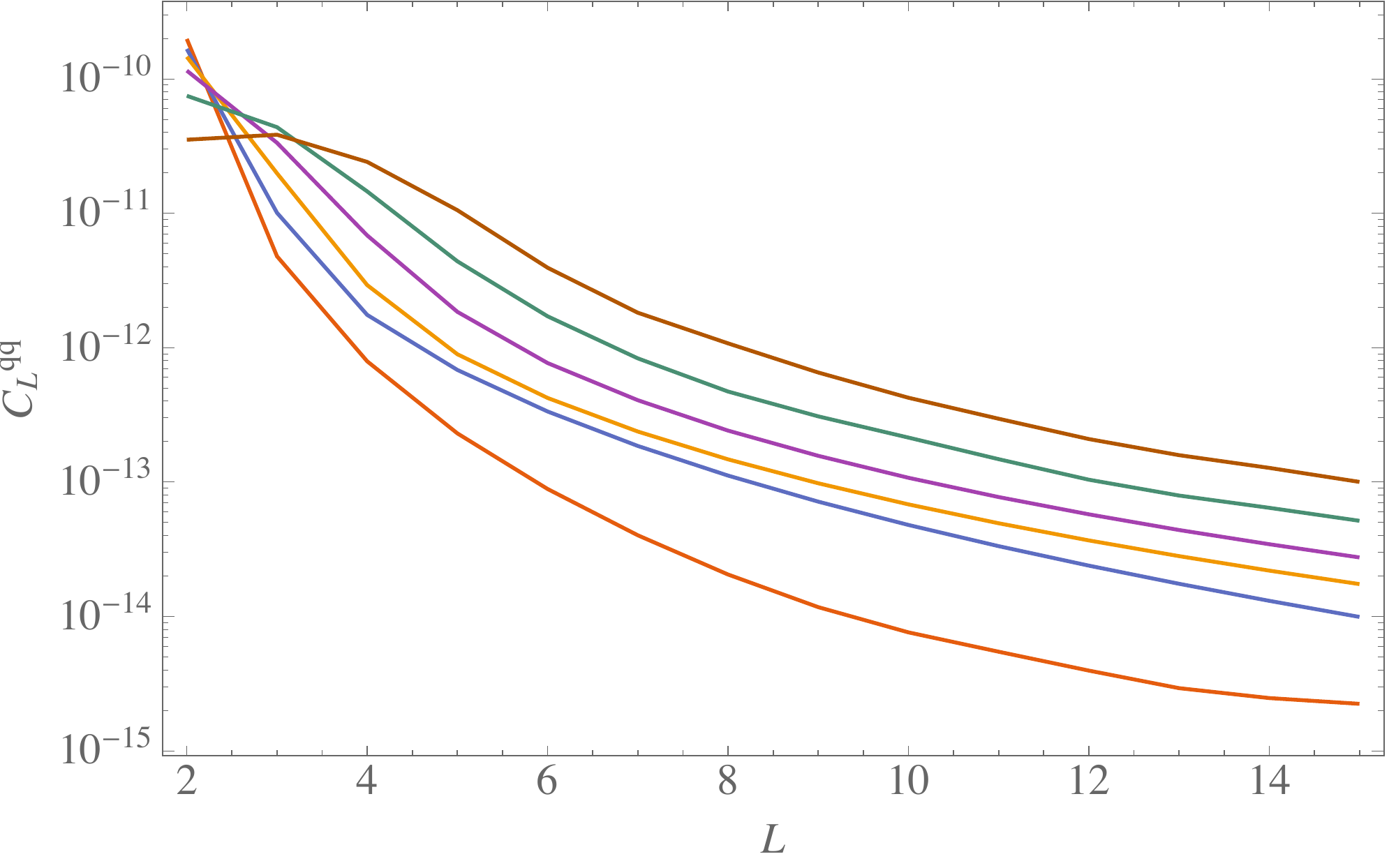}}
		\caption{Left: The effective quadrupole contribution to the signal described in \eqref{eq:clq} at $z = 1$ versus $k_\text{max}$ in units of $H_0$, which refers to the upper bound on the integral. The curves from top to bottom correspond to $L = 2, 3, 4, 5, 6, 7$. Notice that the contribution mainly comes from large scales (small $k$) for low $L$. 
		Right: $C_L^{qq}$ versus $L$ %for $k_\text{max}\sim 278 H_0$ corresponding to $\lambda_\text{min}\sim 100$ Mpc. 
               The different curves, from bottom to top, correspond to increasing values of $\bar{\chi}_e$, taken to be the center of each redshift bin for the six bins described in table~\ref{tab:bins}.}
		\label{fig:clq}
	\end{center}	
\end{figure}

The signal $C_{L}^{X}(\bar{\chi_e})$  depends on the power of the remote quadrupole field
\begin{equation}
	C_L^{qq}(\chi) = \int_0^{k_\text{max}} \frac{k^2dk}{(2\pi)^3} P_\Psi(k)  |\Delta^q_{L}(k, \chi_e)|^2\ , \label{eq:clq}
\end{equation} with $\Delta^q_{L}$ given in equation \eqref{eq:Deltaq}. 
%The upper integration bound, $k_\text{max}$, corresponds to the smallest scale, $\lambda_\text{min}=2\pi/k_\text{max}$, that contributes to the signal. We expect this approximation to work well in the limit of small $L$ where the signal is strongest, and the correlation length for the quadrupole field is not very small compared to the size of the redshift bins.
Figure~\ref{fig:clq} shows $C_L^{qq}(z=1)$ as a function of $k_\text{max}$ (left panel) and as a function of $L$ (right panel) for the six bin scenario defined in table~\ref{tab:bins}. From the left panel, it is clear that the low-$L$ power in the quadrupole field is mainly sensitive to large scales. In the right panel, we see that the power in all redshift bins falls quickly with $L$, with the most dramatic falloff for the low redshift bins. We therefore expect the measurable signal to be dominant at low $L$. \\

\begin{figure}[htbp]
  \begin{center}
			\includegraphics[width=14cm]{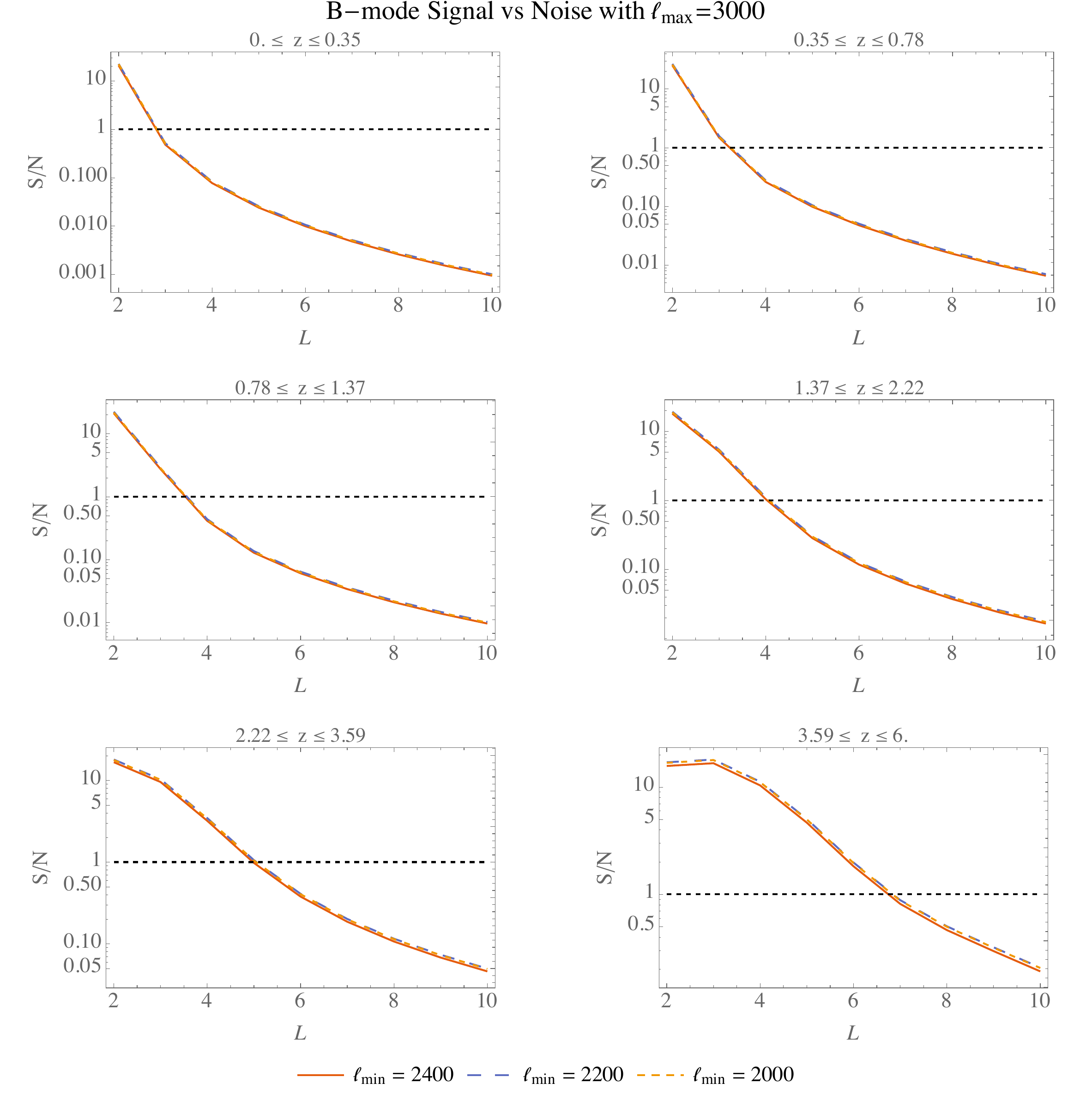}
			\caption{The signal-to-noise ratio of the $B$-mode estimator $\hat{b}_{LM}$, equation \eqref{eq:qSN}, for $\ell_\text{max}=3000$ and six red-shift bins. The optimal value for the filtering scale was chosen numerically, and is given by $\ell_\text{min}=2200$. We find detectability from $L=2$ up to $L=3$ to $L=7$ depending on the red shift bin.}
			 \label{fig:StoNB}
		\end{center} 
\end{figure}

\begin{figure}[htbp]
  \begin{center}
			\includegraphics[width=14cm]{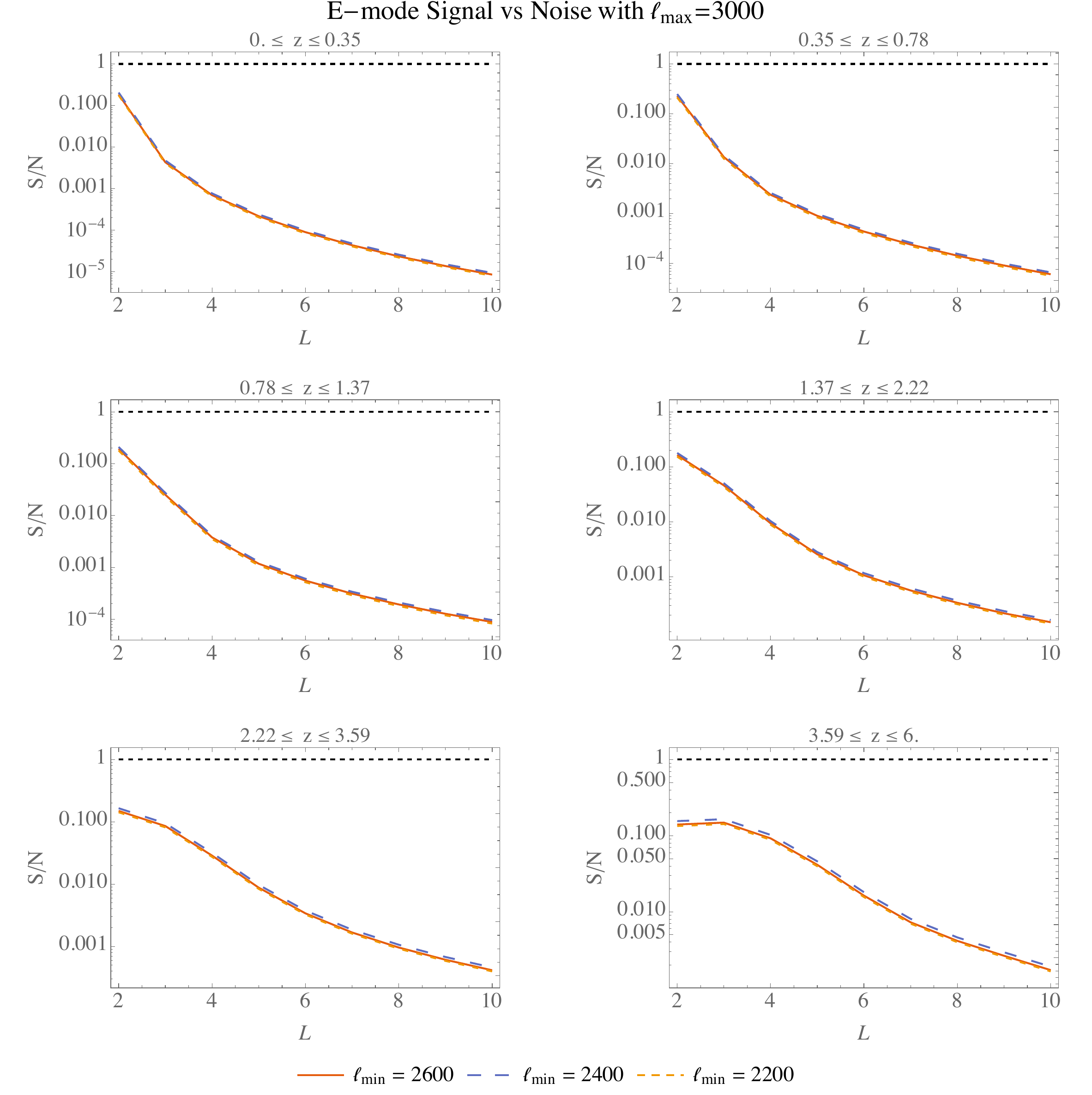}
			\caption{The signal-to-noise ratio of the $E$-mode estimator $\hat{e}_{LM}$, equation \eqref{eq:qSN}, for $\ell_\text{max}=3000$ and six red-shift bins. The optimal value for the filtering scale was chosen numerically, and is given by $\ell_\text{min}=2200$. Unlike for $B$-modes, we do not find detectability for $E$-modes with this value of $\ell_\text{max}=3000$.}
			 \label{fig:StoNE}
		\end{center} 
\end{figure}

Figure~\ref{fig:StoNB} and Figure~\ref{fig:StoNE} show the SNR of the $B$-mode and $E$-mode estimator, as a function of $L$ for the 6 redshift bins described in table~\ref{tab:bins} and for $\ell_\text{max}=3000$. We find that in the $B$-mode case prospects for observation are excellent for this resolution scale, while the noise for $E$-modes is too large. The optimal value of the filtering scale $\ell_\text{min}$ was found numerically by stepping down from the given $\ell_\text{max}$ by $\Delta\ell =50$ and recomputing the sum. We find that the best filtering scale to maximize the SNR is $\ell_\text{min}=2200$. The plots show that for each redshift bin, the signal-to-noise is largest for the lowest power multipoles. There are more detectable modes at high redshift, reflecting the fact that at higher redshift the light cone is large enough to probe variations in the quadrupole field. It is at these high redshifts that we obtain the most information about the quadrupole field.

In conclusion, there is a signal to detect, and both progress in better sensitivity for future CMB experiments and the use of novel techniques such as intensity mapping to probe the angular matter power spectrum at high resolution will improve the detectability of the signal.

\subsection{Information content} \label{sec:modes}

\begin{figure}[htbp]
	\centering
	\includegraphics[width=13cm]{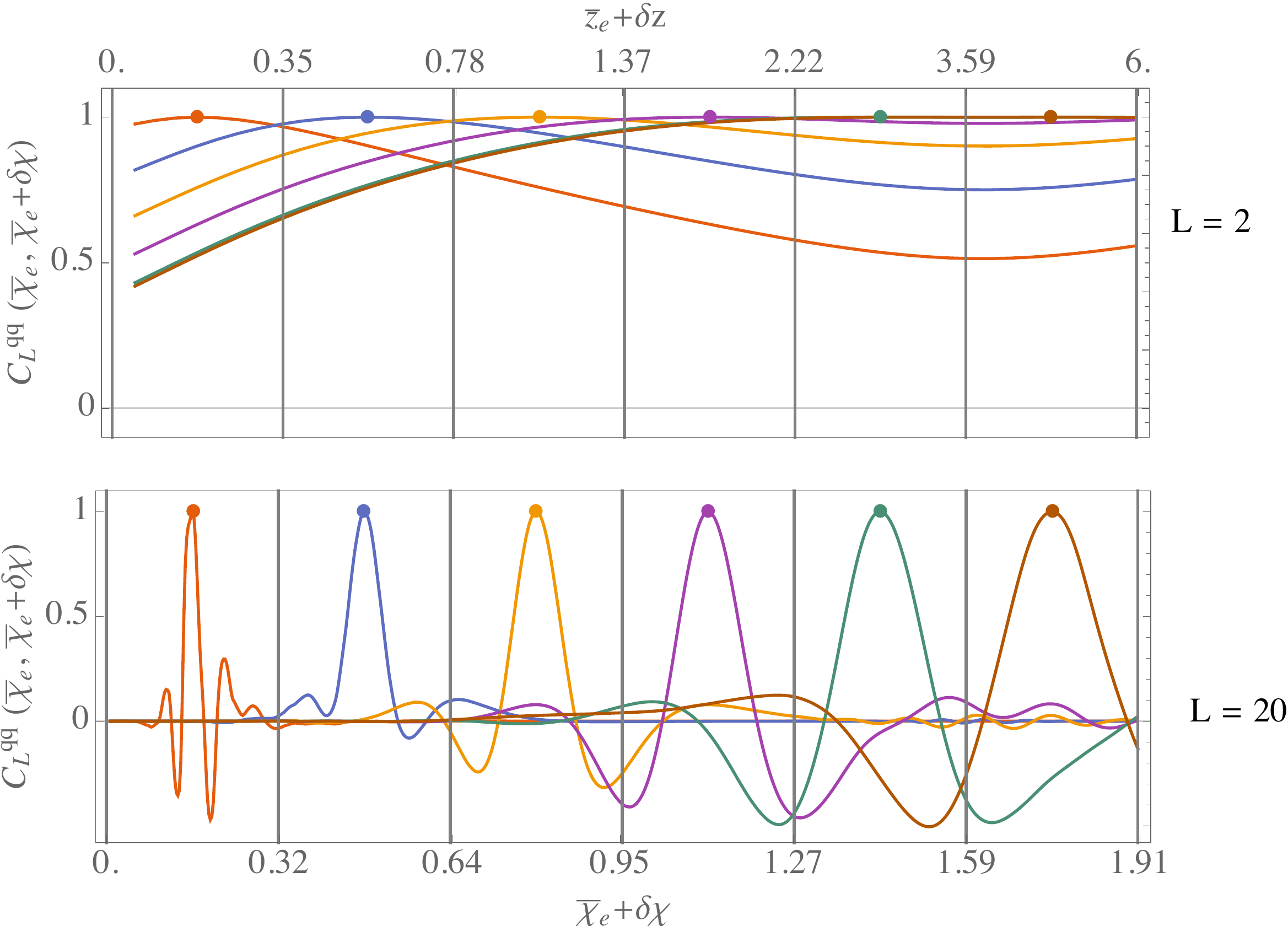}
	\caption{The effective quadrupole correlation in equation~\eqref{eq:Clqq_chi}, normalized as  $C_L^{qq}(\chi_e,\chi'_e)[C_L^{qq}(\chi_e,\chi_e)C_L^{qq}(\chi'_e,\chi'_e)]^{-1/2}$, is plotted at $\chi_e=\bar{\chi}_e,\ \chi'_e = \bar{\chi}_e + \delta\chi$, where $\bar{\chi}_e$ is fixed to be the midpoint of a redshift bin. Here, we consider the six evenly spaced redshift bins in table~\ref{tab:bins} (redshift values displayed along the top axis, comoving distance values displayed along the bottom axis). The different colored curves from left to right fix $\bar{\chi}_e =$ 0.16, 0.48, 0.79, 1.11, 1.43, 1.75, in units of $H_0^{-1}$. The top panel shows $L=2$ for which the correlation is important, and the bottom panel shows $L=20$ for which the correlation is less significant within a given redshift bin.
	}
	\label{fig:Clqq_correlations}
\end{figure}

Having established the in-principle detectability of a signal, how much would we stand to learn from a detection? To address this, we must examine how correlated we expect the $x_{LM} (\bar{\chi}_e)$ to be between redshift bins. This is determined by the correlation function of the quadrupole field, eq.~\eqref{eq:Clqq_chi}. In figure~\ref{fig:Clqq_correlations} we show $C_L^{qq}(\bar{\chi}_e,\bar{\chi}_e+\delta \chi_e)$ centered on each of the six redshift bins of table~\ref{tab:bins} for $L=2$ and $L=20$. Recall that $L=2$ is the lowest non-zero multipole moment, and probes the average quadrupole seen at each redshift. As can be seen from the figure, within $\Lambda$CDM the $L=2$ moment of the quadrupole field is highly correlated between redshift bins, implying that any one redshift bin contains all of the information about the corresponding modes of the primordial curvature perturbation. On the other hand, the $L=20$ moment of the quadrupole field is relatively uncorrelated between redshift bins, implying that each bin can be used to constrain independent modes.

Within a single redshift bin, we can estimate the number of modes as $N \simeq \sum_{L=2}^{L_{\rm max}} 2 L + 1$, where $L_{\rm max}$ is the maximum multipole that can be accessed with a SNR of more than one. For the case discussed above, there are as few as 5 modes in the lowest redshift bins and as many as 60 modes in the highest redshift bin detectable using the $b$-mode estimator. However, as discussed above, these modes at low $L$ are significantly correlated among redshift bins in $\Lambda$CDM, and therefore one does not obtain independent measurements from each redshift bin. 
%A lower bound on the number of independent in principle measurable modes can be obtained from the highest redshift bin, which in the high resolution case described above yields $N > 1443$. 

There are a few important caveats to add to the discussion above. First, while the correlation among redshift bins at low $L$ is not advantageous for constraining a large number of independent modes, it can be used to boost the SNR since the signal in each bin would add coherently. This is equivalent to choosing a different binning scheme for the low $L$ moments of the power asymmetry. Second, the degree of correlation between different redshifts is due not only to the fact that the transfer function eq.~\eqref{eq:Deltaq} depends mainly on long-wavelength modes, but also our assumption of statistical homogeneity. Violating this assumption, as is invoked to explain many of the existing CMB anomalies, there could be less correlation between redshift bins.

\section{Conclusions}
\label{sec:conclusion}

In this paper we have demonstrated the ability of polarized Sunyaev Zel'dovich (pSZ) tomography to measure the remote quadrupole field at high signal-to-noise in an idealized, cosmic variance limit. Anisotropy in the remote quadrupole field at each redshift is encoded in an asymmetry in the cross-power between CMB polarization anisotropies and tracers of large scale structure. The quadrupole field is sensitive to structure on the largest possible scales in the observable Universe. Because early-time, high-energy physics is stretched to ultra-large scales, comparable to, or perhaps much larger than, the size of the observable Universe today, pSZ tomography can potentially make a large impact on our understanding of the early Universe~\footnote{For example, the analogous observable kSZ tomography can in principle improve the constraints on parameters in various early-Universe cosmologies by orders of magnitude over the CMB alone~\cite{Zhang10d,Zhang11b,Zhang:2015uta,Terrana2016}.}. Indeed, the first possible hints of beyond-the-standard-cosmological-model physics may have already been detected in the various anomalies in the large scale primary CMB.

Our primary contribution to previous work on this topic has been to set down the general theoretical formalism for pSZ tomography and define a concrete estimator for the quadrupole field. We have made a number of idealized assumptions that could be improved upon. In particular, we have assumed that electrons trace the dark matter, that the density field is Gaussian, that the contribution to the power asymmetry from lensing can be subtracted, that we can neglect systematics such as non-Gaussian and anisotropic instrumental noise, that foregrounds can be subtracted, that we have data on the full sky, a sub-optimal estimator, and possibly other non-idealities. Nevertheless, this work provides a target for future measurements. As we showed in section~\ref{sec:detection}, the next generation of CMB experiments and galaxy surveys have a chance to detect the first few moments of the quadrupole field in a few redshift bins. This fact motivates a more complete forecast for what might be possible. 

The potential detection of this signal also motivates a more complete assessment of what we might learn about early Universe physics should such an observation be made. Although it is beyond the scope of the present work, the formalism outlined in this paper can be straightforwardly applied to forecasting parameter constraints on any early-Universe model that makes a prediction for the statistical or deterministic properties of large scale modes of the primordial curvature perturbation. Some targets of potential interest include primordial non-Gaussianity, running of the power spectrum, features in the power spectrum, pre-inflationary inhomogeneities, among other scenarios. We hope to perform detailed forecasts in future work. 

\acknowledgments
We thank Gil Holder for important comments and suggestions. MCJ is supported by the National Science and Engineering Research Council through a Discovery grant. AT acknowledges support from the Vanier Canada Graduate Scholarships program. AD is supported by NSF Award PHY-1417385. AD thanks the Perimeter Institute for Theoretical Physics for its hospitality. This research was supported in part by Perimeter Institute for Theoretical Physics. Research at Perimeter Institute is supported by the Government of Canada through the Department of Innovation, Science and Economic Development Canada and by the Province of Ontario through the Ministry of Research, Innovation and Science. Results in this paper were obtained using the HEALPix package \cite{Gorski2005} (\href{http://healpix.sourceforge.net/}{healpix.sourceforge.net}) and the Cosmicpy package (\href{https://cosmicpy.github.io}{cosmicpy.github.io}).

\appendix

\section{Fourier Kernel Calculation}
\label{app:kernel}

In this appendix, we derive the Fourier kernel of the effective quadrupole (eq.~\eqref{eq:initial-density-to-effective-quadrupole},\eqref{eq:kernel}). Each contribution is treated one at a time. Let's begin with the Sachs-Wolfe contribution. Substituting equation~\eqref{eq:psifourier} and \eqref{eqn:SW} into \eqref{eq:effective-quadrupole-def} yields
\begin{equation}
	q_\text{SW}^m ({\bf \hat{n}}_e, \chi_e) =\left( 2D_\Psi(\chi_\text{dec}) -\frac{3}{2} \right) \int_\Omega d^2 {\bf \hat{n}} \int \frac{d^3 k}{(2 \pi)^3} \tilde{\Psi}_i ({\bf k}) e^{i \chi_e {\bf k} \cdot {\bf \hat{n}}_e } e^{i \Delta \chi_\text{dec} {\bf k} \cdot {\bf \hat{n}} }   \ Y^*_{2 m} ({\bf \hat{n}}).
\end{equation}
Next, we employ the identity for the expansion of the exponential,
\begin{equation} \label{eq:expidentity}
	e^{i \Delta \chi {\bf k} \cdot {\bf \hat{n}} } = \sum_{\ell',m'} 4\pi \ i^{\ell'} \ j_{\ell'} (k \Delta \chi) \ Y^*_{\ell'm'}({\bf \hat{k}})  Y_{\ell'm'}({\bf \hat{n}} ),
\end{equation}
resulting in,
\begin{align}
	q_\text{SW}^m ({\bf \hat{n}}_e, \chi_e) =  & \int \frac{d^3 k}{(2 \pi)^3} \tilde{\Psi}_i ({\bf k}) \left( 2D_\Psi(\chi_\text{dec}) -\frac{3}{2} \right) \\
	& \times \int_\Omega d^2 {\bf \hat{n}} \sum_{\ell',m'} 4\pi \ i^{\ell'} j_{\ell'} (k \Delta \chi_\text{dec}) \ Y^*_{\ell'm'}({\bf \hat{k}})  Y_{\ell'm'}({\bf \hat{n}} )  \ Y^*_{2 m} ({\bf \hat{n}}) e^{i \chi_e {\bf k} \cdot {\bf \hat{n}}_e } \no \\
	= & \int \frac{d^3 k}{(2 \pi)^3} \tilde{\Psi}_i ({\bf k})\left[ -4\pi\left( 2D_\Psi(\chi_\text{dec}) -\frac{3}{2} \right) j_{2} (k \Delta \chi_\text{dec}) \right] Y^*_{2m}({\bf \hat{k}}) e^{i \chi_e {\bf k} \cdot {\bf \hat{n}}_e },\label{eq:q_SW}
\end{align}
where we integrated over $\hat{\bf n}$ to obtain $\delta_{\ell' 2}\delta_{m'm}$. 

The steps of the calculation are identical for the integrated Sachs-Wolfe term. We start with $\Theta_\text{ISW}$ in eq.~\eqref{eqn:ISW} to obtain the resulting contribution:
\begin{align}
	q_\text{ISW}^m ({\bf \hat{n}}_e, \chi_e) & = 2 \int_\Omega d^2 {\bf \hat{n}} \int_{a_{\rm
    dec}}^{a_e} da\frac{dD_\Psi}{da} \int \frac{d^3 k}{(2 \pi)^3} \tilde{\Psi}_i ({\bf k}) e^{i \chi_e {\bf k} \cdot {\bf \hat{n}}_e } e^{i \Delta \chi (a) {\bf k} \cdot {\bf \hat{n}} }   \ Y^*_{2 m} ({\bf \hat{n}}) \no \\
    & =  2 \int \frac{d^3 k}{(2 \pi)^3} \tilde{\Psi}_i ({\bf k}) \int_{a_{\rm dec}}^{a_e} da \frac{dD_\Psi}{da} \int_\Omega d^2 {\bf \hat{n}}  \sum_{\ell',m'} 4\pi \ i^{\ell'} j_{\ell'} (k \Delta \chi(a)) \no \\
    & \quad \times Y^*_{\ell'm'}({\bf \hat{k}}) \ Y_{\ell'm'}({\bf \hat{n}}) \ Y^*_{2 m} ({\bf \hat{n}}) \ e^{i \chi_e {\bf k} \cdot {\bf \hat{n}}_e } \no \\
    & = \int \frac{d^3 k}{(2 \pi)^3} \tilde{\Psi}_i ({\bf k}) \left[ -8\pi \int_{a_{\rm dec}}^{a_e} da \frac{dD_\Psi}{da}  \ j_{2} (k \Delta \chi (a)) \right] Y^*_{2m}({\bf \hat{k}}) e^{i \chi_e {\bf k} \cdot {\bf \hat{n}}_e }. \label{eq:q_ISW}
\end{align}
Both the SW and ISW contributions are sensitive mainly to large scale potential fluctuations, as shown in figure~\ref{fig:kernel}.

The derivation of the Doppler kernel requires more work. However, the contribution from the second term in eq.~\eqref{eq:thetaDopp} vanishes. To show this, we substitute the second term of $\Theta_\text{Doppler}$ into the effective quadrupole, yielding,
\begin{equation}
	 -D_v(\chi_e) \int_\Omega d^2 {\bf \hat{n}} \int \frac{d^3 k}{(2 \pi)^3} \ i k \tilde{\Psi}_i ({\bf k}) (\hat{\bf n}\cdot \hat{\bf k}) e^{i \chi_e {\bf k} \cdot {\bf \hat{n}}_e } \ Y^*_{2 m} ({\bf \hat{n}}).
\end{equation}
It is equivalent to write  $(\hat{\bf n}\cdot \hat{\bf k})$ as $\mathcal{P}_1 (\hat{\bf n}\cdot \hat{\bf k})$ and then expand the Legendre polynomial in terms of spherical harmonics using
 \begin{equation} \label{eq:legendre}
 	\mathcal{P}_\ell ({\bf \hat{x}}\cdot{\bf \hat{x}}')=\frac{4\pi}{2\ell +1} \sum_{m=-\ell }^\ell Y^*_{\ell m}({\bf \hat{x}})\ Y_{\ell m}({\bf \hat{x}}').
 \end{equation}
Doing so results in
\begin{equation}
	-D_v(\chi_e) \int \frac{d^3 k}{(2 \pi)^3} \ i k \tilde{\Psi}_i ({\bf k}) e^{i \chi_e {\bf k} \cdot {\bf \hat{n}}_e }  \int_\Omega d^2 {\bf \hat{n}}\ \frac{4\pi}{3} \sum_{m'=-1 }^1 Y^*_{1 m'}({\bf \hat{k}})\ Y_{1 m'}({\bf \hat{n}})  Y^*_{2 m} ({\bf \hat{n}}) =0 ,
\end{equation}
which vanishes upon integration over $\hat{\bf n}$ due to the orthogonality of the spherical harmonics. The first term of eq.~\eqref{eq:thetaDopp} gives a non-zero contribution to the effective quadrupole. The calculation proceeds similarly, except for the additional factor of $e^{i \Delta \chi_\text{dec} {\bf k} \cdot {\bf \hat{n}} }$ which needs to be expanded using \eqref{eq:expidentity}:
\begin{align}
	q_\text{Doppler}^m ({\bf \hat{n}}_e, \chi_e) & = D_v(\chi_\text{dec}) \int_\Omega d^2 {\bf \hat{n}} \int \frac{d^3 k}{(2 \pi)^3} \ i k \tilde{\Psi}_i ({\bf k}) (\hat{\bf n}\cdot \hat{\bf k}) e^{i \chi_e {\bf k} \cdot {\bf \hat{n}}_e }e^{i \Delta \chi_\text{dec} {\bf k} \cdot {\bf \hat{n}} } \ Y^*_{2 m} ({\bf \hat{n}}) \no \\
\begin{split}
	& = D_v(\chi_\text{dec}) \int_\Omega d^2 {\bf \hat{n}} \int \frac{d^3 k}{(2 \pi)^3} \ i k \tilde{\Psi}_i ({\bf k}) \ \frac{4\pi}{3} \sum_{m'=-1 }^1 Y^*_{1 m'}({\bf \hat{k}})\ Y_{1 m'}({\bf \hat{n}}) \\
	& \quad \times \sum_{\ell'',m''}  4\pi \ i^{\ell''} j_{\ell''} (k \Delta \chi(a)) \ Y^*_{\ell''m''}({\bf \hat{k}})  Y_{\ell''m''}({\bf \hat{n}} ) e^{i \chi_e {\bf k} \cdot {\bf \hat{n}}_e }\ Y^*_{2 m} ({\bf \hat{n}}).
\end{split}
\end{align}
In the above expression, the integral over $\hat{\bf n}$ is a triple product of spherical harmonics. For this we can apply the general identity in terms of Wigner 3-$j$ symbols,
\begin{equation}
\begin{split}
\int_\Omega d^2 {\bf \hat{n}} \ {}_{s_1}Y_{\ell_1 m_1} ({\bf \hat{n}}) {}_{s_2}Y_{\ell_2 m_2} ({\bf \hat{n}}) {}_{s_3}Y_{\ell_3 m_3} ({\bf \hat{n}}) & = \sqrt{ \frac{(2\ell_1 +1) (2 \ell_2+1)(2\ell_3+1)}{4\pi} } \\
& \quad \times \left(\begin{array}{ccc} \ell_1 & \ell_2 & \ell_3 \\ m_1 & m_2 & m_3 \end{array} \right) \left(\begin{array}{ccc} \ell_1 & \ell_2 & \ell_3 \\ -s_1 & -s_2 & -s_3 \end{array} \right)
\end{split}
\label{eq:identity-three-spherical-harmonics}
\end{equation}
with spin weights $s_1=s_2=s_3=0$.  There are also two spherical harmonics with argument  $\hat{\bf k}$ that can be expressed as a single spherical harmonic using another identity,
\begin{equation}
\begin{split}
	{}_{s_1}Y_{\ell_1 m_1} ({\bf \hat{n}}) \ {}_{s_{2}} Y_{\ell_{2} m_{2}}({\bf \hat{n}}) = \sum_{S,L,M} & (-1)^{\ell_1+\ell_{2}+L} \sqrt{ \frac{(2\ell_1+1) (2\ell_{2}+1)(2L+1)}{4\pi} } \\
	& \quad \times \left(\begin{array}{ccc} \ell_1 & \ell_{2} & L \\ m_1 & m_{2} & M \end{array} \right) \left(\begin{array}{ccc} \ell_1 & \ell_2 & L \\ s_1 & s_{2} & S \end{array} \right) {}_{S}Y_{LM}^{*}({\bf \hat{n}}).
	\end{split} \label{eq:identity-two-spherical-harmonics}
\end{equation}
Putting all of this together yields
\begin{align}
	q_\text{Doppler}^m ({\bf \hat{n}}_e, \chi_e) & = D_v(\chi_\text{dec}) \int \frac{d^3 k}{(2 \pi)^3} \ i k \tilde{\Psi}_i ({\bf k}) \sum_{\substack{L,M,m', \\ \ell'',m''}} \sqrt{\frac{(5)(3)(2\ell''+1)}{4\pi}} \no \\
	& \quad \times \sqrt{\frac{(2L+1)(3)(2\ell''+1)}{4\pi}} \frac{(4\pi)^2}{3}  i^{\ell''} j_{\ell''}(k \Delta \chi_\text{dec}) (-1)^{\ell''+L+1} \label{eq:doppler_4wigs} \\
	& \quad \times \left(\begin{array}{ccc} 1 & \ell'' & L \\ m' & m'' & M \end{array} \right) \left(\begin{array}{ccc} 2 & 1 & \ell'' \\ m & m' & m'' \end{array} \right) \left(\begin{array}{ccc} 2 & 1 & \ell'' \\ 0 & 0 & 0 \end{array} \right) \left(\begin{array}{ccc} 1 & \ell'' & L \\ 0 & 0 & 0 \end{array} \right) Y^*_{LM}({\bf \hat{k}}) e^{i \chi_e {\bf k} \cdot {\bf \hat{n}}_e }.\no
\end{align}
Fortunately, the sum over $m''$ and $m'$ simplify the expression drastically because of the orthogonality relation,
\begin{equation} \label{eq:wigner_orthogonal}
	\sum_{m_1,m_2} \left(\begin{array}{ccc} \ell_1 & \ell_{2} & L \\ m_1 & m_{2} & M \end{array} \right) \left(\begin{array}{ccc} \ell_1 & \ell_2 & L' \\ m_1 & m_{2} & M' \end{array} \right) = \frac{\delta_{LL'}\delta_{MM'}}{2L+1}.
\end{equation}
Using the invariance of the Wigner 3-$j$ symbols under even permutations of its columns, we can apply this relation to the first two 3-$j$ symbols in eq.~\eqref{eq:doppler_4wigs}, it follows that,
\begin{equation}
\begin{split}
	q_\text{Doppler}^m ({\bf \hat{n}}_e, \chi_e) & = D_v(\chi_\text{dec}) \int \frac{d^3 k}{(2 \pi)^3} \ i k \tilde{\Psi}_i ({\bf k}) \sum_{\ell''} 4\pi i^{\ell''} j_{\ell''}(k \Delta \chi_\text{dec}) \\
	& \quad \times (-1)^{\ell''+1} (2\ell''+1) \left(\begin{array}{ccc} 2 & 1 & \ell'' \\ 0 & 0 & 0 \end{array} \right)^2 Y^*_{2m}({\bf \hat{k}}) e^{i \chi_e {\bf k} \cdot {\bf \hat{n}}_e }.
\end{split}
\end{equation}
The remaining 3-$j$ symbol is only non-zero for $\ell'' =$ 1 and 3, resulting in the final expression for the Doppler contribution,
\begin{align}
	q_\text{Doppler}^m ({\bf \hat{n}}_e, \chi_e) = & \int \frac{d^3 k}{(2 \pi)^3} \tilde{\Psi}_i ({\bf k})  \times Y^*_{2m}({\bf \hat{k}}) e^{i \chi_e {\bf k} \cdot {\bf \hat{n}}_e }  \no \\
	& \times \left[ \frac{4\pi}{5} k D_v(\chi_\text{dec}) \left( 3j_3(k\Delta\chi_\text{dec}) - 2j_1(k\Delta\chi_\text{dec}) \right) \right]. \label{eq:q_doppler}
\end{align}
As illustrated in figure~\ref{fig:kernel}, the Doppler term dominates over SW and ISW contributions on small scales $k \gtrsim 20 H_0$. 

The full expression for the effective quadrupole is a sum of the contributions in~\eqref{eq:q_SW}, \eqref{eq:q_ISW} and~\eqref{eq:q_doppler}:
\begin{equation}
q^{m}_{\rm eff} ({\bf \hat{n}}_e, \chi_e) = \int \frac{d^3k}{(2 \pi)^3}  \tilde{\Psi}_i({\bf k}) T(k)  \left[ \mathcal{G}_{\rm SW} + \mathcal{G}_{\rm ISW} + \mathcal{G}_{\rm Doppler } \right] Y_{2 m}^*({\bf \hat{k}}) \ e^{i \chi_e {\bf k} \cdot {\bf \hat{n}}_e},
\end{equation}
where we have added the transfer function $T(k)$ which has the form given in~\eqref{eq:Tk}, and the kernels $\mathcal{G}_{\rm SW}$, $\mathcal{G}_{\rm ISW}$ and $\mathcal{G}_{\rm Doppler }$ are given by,
\begin{align} 
	\mathcal{G}_{\rm SW}(k,\chi_e) & = -4\pi \left( 2D_\Psi(\chi_{\rm dec}) -\frac{3}{2} \right) j_{2} (k \Delta \chi_\text{dec}), \no \\
	\mathcal{G}_{\rm ISW}(k,\chi_e) & = -8\pi \int_{a_{\rm dec}}^{a_e} da \frac{dD_\Psi}{da}  \ j_{2} (k \Delta \chi (a)), \no \\
	\mathcal{G}_{\rm Doppler }(k,\chi_e) & = \frac{4\pi}{5} k D_v(\chi_\text{dec}) \left[ 3j_3(k\Delta\chi_\text{dec}) - 2j_1(k\Delta\chi_\text{dec}) \right]. 
\end{align}

\section{Multipole coefficients for the total effective quadrupole}
\label{app:almq}

Here we compute $a_{\ell m}^q(\chi_e)$ starting from \eqref{eq:almqdef}. Inserting the total effective quadrupole \eqref{eq:qeffeff} and the expression for $q_\text{eff}^m$ from \eqref{eq:initial-density-to-effective-quadrupole}, we have 
\begin{equation}
\begin{split}
	a_{\ell m}^q(\chi_e) & = \int_\Omega d^2 {\bf \hat{n}}_e \sum_{m'=-2}^2 \int \frac{d^3 k}{(2\pi)^3}  \tilde{\Psi}_i({\bf k}) T(k)  \left[ \mathcal{G}_{\rm SW} + \mathcal{G}_{\rm ISW} +  \mathcal{G}_{\rm Doppler} \right] \\
	& \quad \times Y_{2 m'}^*({\bf \hat{k}}) \ e^{i \chi_e {\bf k} \cdot {\bf \hat{n}}_e} \left._{\pm 2}Y_{2 m'}\right. ({\bf \hat{n}}_e) \left._{\pm 2}Y^*_{\ell m}\right. ({\bf \hat{n}}_e) \ .
\end{split}
\end{equation}
When we expand the exponential with the exponential identity \eqref{eq:expidentity} introducing new multipole parameters $L,\ M$, there will be five spherical harmonics, three with argument ${\bf \hat{n}}_e$: $\left._{\pm 2}Y_{2 m'}\right. ({\bf \hat{n}}_e), \left._{\pm 2}Y^*_{\ell m}\right. ({\bf \hat{n}}_e)$ and $Y_{LM}({\bf \hat{n}}_e)$, and two with argument ${\bf \hat{k}}$: $Y_{2 m'}^*({\bf \hat{k}})$ and $Y_{LM}^*({\bf \hat{k}})$. The first three can be handled by the triple-product spin-weighted spherical harmonic integral identity in eq.~\eqref{eq:identity-three-spherical-harmonics}. Using this to integrate the spherical harmonics with argument ${\bf \hat{n}}_e$ gives
\begin{equation} 
\begin{split}
\int_\Omega d^2 {\bf \hat{n}} \ \left._{\pm 2}Y_{2 m'}\right. ({\bf \hat{n}}_e) \left._{\pm 2}Y^*_{\ell m}\right. ({\bf \hat{n}}_e) Y_{LM}({\bf \hat{n}}_e) & = (-1)^m\sqrt{ \frac{(5)(2 \ell+1)(2L+1)}{4\pi} } \\
& \quad \times \left(\begin{array}{ccc} \ell & 2 & L \\ -m & m' & M \end{array} \right) \left(\begin{array}{ccc} \ell & 2 & L \\ \pm2 & \mp2 & 0 \end{array} \right).
\end{split}\label{eq:YYY}
\end{equation}
For the remaining two spherical harmonics with argument $\hat{\bf k}$, we can use the identity in eq.~\eqref{eq:identity-two-spherical-harmonics} to express them as just one spherical harmonic, which results in
\begin{equation}
\begin{split}
	Y_{LM}^{*}({\bf \hat{k}}) Y^{*}_{2m^{\prime}}({\bf \hat{k}}) & = (-1)^{M+m^{\prime}} \sum_{L^{\prime} M^{\prime}} (-1)^{L+L^{\prime}} \sqrt{ \frac{5(2L+1)(2L^{\prime}+1)}{4\pi} } \\
	& \quad \times \left( \begin{array}{ccc} L & 2 & L^{\prime} \\ -M & -m^{\prime} & M^{\prime} \end{array} \right) \left( \begin{array}{ccc} L & 2 & L^{\prime} \\ 0 & 0 & 0 \end{array} \right) Y^{*}_{L^{\prime}M^{\prime}}({\bf \hat{k}}).
\end{split} \label{eq:YY}
\end{equation}
Notice that when we combine the results of equations \eqref{eq:YYY} and \eqref{eq:YY} there are four Wigner 3-$j$ symbols. However, there is a nice simplification when we perform the sums over $m'$ and $M$ due to the relation~\eqref{eq:wigner_orthogonal},
\begin{align}
	\sum_{M,m^{\prime}} (-1)^{M+m^{\prime}+m} \left( \begin{array}{ccc} \ell & 2 & L \\ -m & m^{\prime} & M \end{array} \right) \left( \begin{array}{ccc} L & 2 & L^{\prime} \\ -M & -m^{\prime} & M^{\prime} \end{array} \right) & = \sum_{M,m^{\prime}} \left( \begin{array}{ccc} L & 2 & \ell \\ M & m^{\prime} & -m \end{array} \right) \left( \begin{array}{ccc} L & 2 & L^{\prime} \\ M & m^{\prime} & -M^{\prime} \end{array} \right) \no \\
	 & = \frac{\delta_{\ell L^{\prime}} \delta_{m M^{\prime}}}{2 \ell +1},
\end{align}
where we used the selection rule of the 3-$j$ symbols $M+m^{\prime}-m=0$. Then, owing to the fact that $\left( \begin{array}{ccc} \ell & 2 & L \\ 0 & 0 & 0 \end{array} \right)$ vanishes if $\ell + 2 + L$ is odd, and
\begin{equation}
	\left( \begin{array}{ccc} \ell & 2 & L \\ 2 & -2 & 0 \end{array} \right) = (-1)^{\ell +L} 	\left( \begin{array}{ccc} \ell & 2 & L \\ -2 & 2 & 0 \end{array} \right),
\end{equation}
we see that both contributions from $\pm 2$ are equal.
The result thus far reads,
\begin{align}
	a_{\ell m}^q(\chi_e) = &\int \frac{d^3 k}{(2\pi)^3} \tilde{\Psi}_i({\bf k})\ T(k) \left[ \mathcal{G}_{\rm SW} + \mathcal{G}_{\rm ISW} +  \mathcal{G}_{\rm Doppler} \right] \no \\
	& \times\sum_L 5i^L(2L+1) \left(\begin{array}{ccc} \ell & 2 & L \\ 0 & 0 & 0 \end{array} \right) \left(\begin{array}{ccc} \ell & 2 & L \\ \pm2 & \mp2 & 0 \end{array} \right)  j_L(k\chi_e) \ Y_{\ell m}^*({\bf \hat{k}}).
\end{align}

This expression can be further simplified. Indeed, for the 3-$j$ symbols to be non-zero, the selection rule $\mid\ell-2\mid \leq L \leq \ell + 2$ needs to be satisfied. This means that for all $\ell \geq 2$, only the terms $L=\ell-2,\ \ell,\ \ell+2$ will contribute. The 3-$j$ symbols can then be expressed in each case as:
\begin{align}
	\left(\begin{array}{ccc} \ell & 2 & \ell-2 \\ \pm 2 & \mp 2 & 0 \end{array}\right) \left(\begin{array}{ccc} \ell & 2 & \ell-2 \\ 0 & 0 & 0 \end{array}\right) & = \sqrt{\frac{3}{8}} \sqrt{\frac{(\ell+2)!}{(\ell-2)!}} \frac{1}{(2\ell-3)(2\ell+1)(2\ell-1)},\\
	\left(\begin{array}{ccc} \ell & 2 & \ell \\ \pm 2 & \mp 2 & 0 \end{array}\right) \left(\begin{array}{ccc} \ell & 2 & \ell \\ 0 & 0 & 0 \end{array}\right) & = -\sqrt{\frac{3}{2}} \sqrt{\frac{(\ell+2)!}{(\ell-2)!}} \frac{1}{(2\ell+1)(2\ell-1)(2\ell+3)},\\
	\left(\begin{array}{ccc} \ell & 2 & \ell+2 \\ \pm 2 & \mp 2 & 0 \end{array}\right) \left(\begin{array}{ccc} \ell & 2 & \ell+2 \\ 0 & 0 & 0 \end{array}\right) & = \sqrt{\frac{3}{8}} \sqrt{\frac{(\ell+2)!}{(\ell-2)!}} \frac{1}{(2\ell+5)(2\ell+3)(2\ell+1)}.
\end{align}
Therefore, we can write the sum over $L$ as,
\begin{align}
\begin{split}
	\sum_L & i^L 5(2L+1) j_L(\chi_e k) \left(\begin{array}{ccc} \ell & 2 & L \\ \pm 2 & \mp 2 & 0 \end{array}\right) \left(\begin{array}{ccc} \ell & 2 & L \\ 0 & 0 & 0 \end{array}\right) \\
	& = 5 i^{\ell} \sqrt{ \frac{(\ell+2)!}{(\ell-2)!}} \sqrt{\frac{3}{8}} \left[\frac{-(2\ell-3) j_{\ell-2}}{(2\ell-3)(2\ell-1)(2\ell+1)} \right. \\
	& \quad \left. - 2\frac{(2\ell+1) j_{\ell}}{(2\ell-1)(2\ell+1)(2\ell+3)} + \frac{-(2\ell+5)j_{\ell+2}}{(2\ell+1)(2\ell+3)(2\ell+5)} \right]
\end{split} \no \\
	&= -5 i^{\ell} \sqrt{\frac{3}{8}} \sqrt{\frac{(\ell+2)!}{(\ell-2)!}} \frac{j_{\ell}(\chi_e k)}{(k \chi_e)^2},
\end{align}
where the last line uses recursion relations for the spherical Bessel functions \cite{peter2013}. We can now construct the final expression,
\begin{equation}
	a_{\ell m}^q(\chi_e) = \int \frac{d^3k}{(2\pi)^3} \Delta_\ell^q(k,\chi_e)\ \tilde{\Psi}_i({\bf k})\ Y^*_{\ell m}({\bf \hat{k}}) \ ,
\end{equation}
where the transfer function for the quadrupole is 
\begin{equation*}
	\Delta_\ell^q(k,\chi_e) = -5 i^\ell \sqrt{\frac{3}{8}}\sqrt{\frac{(\ell+2)!}{(\ell-2)!}} \frac{j_\ell(k\chi_e)}{(k\chi_e)^2}T(k) \left[ \mathcal{G}_{\rm SW}(k,\chi_e) + \mathcal{G}_{\rm ISW}(k,\chi_e) +  \mathcal{G}_{\rm Doppler}(k,\chi_e) \right].
\end{equation*}

\section{Contributions to the pSZ power spectrum}
\label{sec:contributions-psz-ps}

In this appendix, we compute the contributions to the pSZ $E$-mode power spectrum Eq.~\ref{eq:ClEpSZqq} and \ref{eq:ClEpSZ1} as well as the contributions to the pSZ $B$-mode power spectrum \ref{eq:ClBpSZ1}.

\subsection{$C_\ell^{EE,{\rm pSZ}, (0)}$}
\label{sec:contributions-from-qq}

Starting from Eq.~\ref{eq:pSZE0}, the contribution to the pSZ power from the homogeneous component of the electron density field is:
\begin{align}
	C_\ell^{EE,\text{pSZ}\ (0)} & = \frac{6\sigma_T^2}{100} \int d\chi_e a_e \bar{n}_e \int d\chi'_e a'_e \bar{n}'_e \ C_L^{qq}(\chi_e,\chi'_e) \no \\
	& = \frac{6\sigma_T^2}{100} \int \frac{dk\ k^2}{(2\pi)^3} P_\Psi(k) \int d\chi_e a_e \bar{n}_e \Delta_\ell^q(k,\chi_e) \int d\chi'_e a'_e \bar{n}'_e \Delta_\ell^{q,*}(k,\chi'_e) \no \\
	& \simeq \frac{6\sigma_T^2}{100} \int \frac{d\chi}{\chi^2} P_\Psi(k) \ a_e^2(\chi)\ \bar{n}_e^2(\chi) \left(\frac{5}{4\pi} \sqrt{\frac{3}{8}}\sqrt{\frac{(\ell+2)!}{(\ell-2)!}}\right)^2 \no \\
	& \quad \times \left[ \frac{T(k)}{(k\chi)^2} \left[ \mathcal{G}_{\rm SW}(k,\chi) + \mathcal{G}_{\rm ISW}(k,\chi) +  \mathcal{G}_{\rm Doppler}(k,\chi) \right] \right]^2 \Bigg|_{k \rightarrow (\ell+1/2)/\chi} ,
\end{align}
where the integral runs from $\chi=0$ to reionization. We expect this to be a purely large scale contribution to the power spectrum, and indeed, figure~\ref{fig:ClEEpSZ} shows that $C_\ell^{EE,\text{pSZ}\ (0)}$ is only significant at $\ell < 10$.

\subsection{$C_\ell^{EE,{\rm pSZ}, (1)}$ and $C_\ell^{BB,{\rm pSZ}, (1)}$}
\label{sec:contributions-from-qq-dd}

Let's now consider the contribution to the pSZ power from the inhomogeneous distribution of electrons, $C_\ell^{EE,\text{pSZ}, (1)}$. Starting from Eq.~\ref{eq:pSZE1}, we have for the correlation function:
\begin{equation}
\begin{split}
	\left< E_{\ell m}E^*_{\ell' m'} \right>^{\text{pSZ},(1)} =& \frac{6\sigma_T^2}{100} \sum_{L,M} \sum_{L',M'} \int d\chi_e a_e \bar{n}_e \int d\chi'_e a'_e \bar{n}'_e \frac{1}{4} \left( 1 + (-1)^{\ell+L+L'} \right) \left( 1 + (-1)^{\ell'+L+L'} \right) \no \\
	& \times \frac{(2L+1) (2L'+1) \sqrt{(2\ell+1) (2 \ell'+1)}}{4\pi} C_L^{qq}(\chi_e,\chi'_e)\ C_{L'}^{\delta\delta}(\chi_e,\chi'_e) \no \\
	& \times \left(\begin{array}{ccc} \ell & L & L' \\ -m & M & M' \end{array} \right) \left(\begin{array}{ccc} \ell' & L & L' \\ -m' & M & M' \end{array} \right) \left(\begin{array}{ccc} \ell'& L & L' \\ -2 & 2 & 0 \end{array} \right) \left(\begin{array}{ccc} \ell & L & L' \\ -2 & 2 & 0 \end{array} \right).
\end{split}
\end{equation}
Collecting the sum over $M$ and $M'$ allows us to simplify two of these 3-$j$ symbols using the orthogonality relation~\eqref{eq:wigner_orthogonal},
\begin{align}
	\sum_{M,M'} & \left(\begin{array}{ccc} \ell & L & L' \\ -m & M & M' \end{array} \right)  \left(\begin{array}{ccc} \ell' & L & L' \\ -m' & M & M' \end{array} \right) \no \\
	& =\sum_{M,M'} (-1)^{\ell+L+L'} \left(\begin{array}{ccc} L & L' & \ell \\ M & M' & -m \end{array} \right)  (-1)^{\ell'+L+L'} \left(\begin{array}{ccc} L & L' & \ell' \\ M & M' & -m' \end{array} \right) \no \\
	& = \frac{\delta_{\ell\ell'}\delta_{mm'}}{2\ell+1}.
\end{align}

Putting this together we have,
\begin{equation}
\begin{split}
	C_\ell^{EE,\text{pSZ}, (1)}  = & \frac{6\sigma_T^2}{100} \sum_{L,L'}  \int d\chi_e a_e \bar{n}_e \int d\chi'_e a'_e \bar{n}'_e \ C_L^{qq}(\chi_e,\chi'_e)\ C_{L'}^{\delta\delta}(\chi_e,\chi'_e) \\ 
	& \times \frac{(2L+1)(2L'+1)}{4\pi} \frac{1}{4} \left( 1 + (-1)^{\ell+L+L'} \right)^2 \left(\begin{array}{ccc} \ell & L & L' \\ \mp 2 & \pm 2 & 0 \end{array} \right)^2 .
\end{split}
\end{equation}
We can make this calculation more tractable by putting it in a form that allows us to use the Limber approximation. To do this, we use the expression for $C_{L'}^{\delta\delta}(\chi_e,\chi'_e)$ so that the integrals become
\begin{align}
	f_{L,L'} & \equiv \int d\chi_e a_e \bar{n}_e \int d\chi'_e a'_e \bar{n}'_e \ C_L^{qq}(\chi_e,\chi'_e) \int \frac{dk\ k^2}{(2\pi)^3} \ 4\pi\ j_{L'}(k\chi_e) \no \\
	& \qquad \times \sqrt{P_\delta(k,\chi_e)}\ 4\pi\ j_{L'}(k\chi'_e) \sqrt{P_\delta(k,\chi'_e)} \no \\
	& = \int dk \frac{2k^2}{\pi} \int d\chi_e a_e \bar{n}_e  \  j_{L'}(k\chi_e) \sqrt{P_\delta(k,\chi_e)} \no \\
	& \qquad \times \int d\chi'_e a'_e \bar{n}'_e \ C_L^{qq}(\chi_e,\chi'_e)\  j_{L'}(k\chi'_e) \sqrt{P_\delta(k,\chi'_e)} \no \\
	& \simeq \int \frac{dk}{L'+1/2}\ C_L^{qq}(\chi)\ a^2_e(\chi)\  \bar{n}^2_e(\chi)\ P_\delta(k,\chi)\ \Big|_{\chi \rightarrow (L'+1/2)/k} \no \\
	& \simeq \int \frac{d\chi}{\chi^2} \ C_L^{qq}(\chi) \ a^2_e(\chi)\  \bar{n}^2_e(\chi)\ P_\delta\left( \frac{L' + 1/2}{\chi},\chi \right) .
\end{align}
We can compute the power spectrum by first calculating $C_L^{qq}(\chi)$, then evaluating the Limber approximation to find $f_{L,L'}$, and summing everything together over $L$ and $L'$:
\begin{equation}\label{eq:ClEpSZapp}
	C_\ell^{EE,\text{pSZ}, (1)} = \frac{6\sigma_T^2}{100} \sum_{L,L'} \frac{(2L+1)(2L'+1)}{4\pi} \frac{1}{4} \left( 1 + (-1)^{\ell+L+L'} \right)^2 \left(\begin{array}{ccc} \ell & L & L' \\ \mp 2 & \pm 2 & 0 \end{array} \right)^2 f_{L,L'}.
\end{equation}

The computation for the $B$-mode power spectrum from pSZ proceeds analogously, resulting in 
\begin{equation}\label{eq:ClBpSZapp}
	C_\ell^{BB,\text{pSZ}, (1)} = \frac{6\sigma_T^2}{100} \sum_{L,L'} \frac{(2L+1)(2L'+1)}{4\pi} \frac{1}{4} \left( 1 - (-1)^{\ell+L+L'} \right)^2 \left(\begin{array}{ccc} \ell & L & L' \\ \mp 2 & \pm 2 & 0 \end{array} \right)^2 f_{L,L'}.
\end{equation}

\section{Lensing Potential}
\label{sec:lensingpotential}

\begin{figure}[htbp]
  \begin{center}
			\includegraphics[width=10cm]{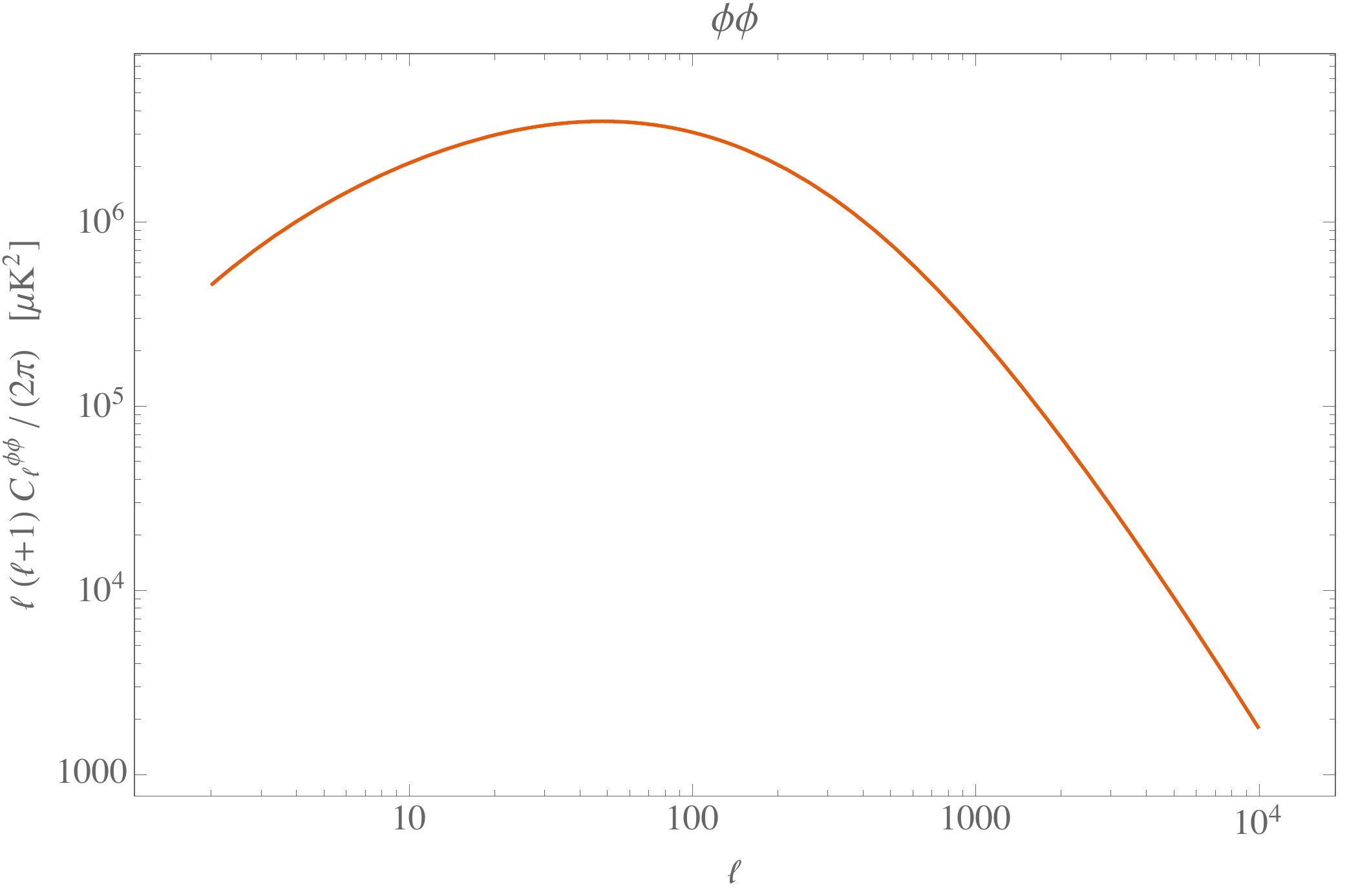}
			\caption{The lensing potential power spectrum, computed using equation~\eqref{eq:Clphiphi}}
			 \label{fig:Clphiphi}
		\end{center} 
\end{figure}

The lensing potential $\phi$ is defined as
\begin{equation}
	\phi ({\bf \hat{n}}) = -2 \int_{0}^{\chi_{\rm dec}} d \chi \ \frac{\chi_{\rm dec} - \chi}{\chi \chi_{\rm dec}} \Psi(\chi {\bf \hat{n}}, \chi).
\end{equation}
In harmonic space,
\begin{align}
	\phi ({\bf \hat{n}}) & = -2 \int_{0}^{\chi_{\rm dec}} d \chi \ \frac{\chi_{\rm dec} - \chi}{\chi \chi_{\rm dec}}  \int \frac{d^3k}{(2 \pi)^3} \tilde{\Psi}({\bf k}, \chi) e^{i \chi {\bf k} \cdot \hat{n}} \no \\
	& = -2 \int \frac{d^3k}{(2 \pi)^3} \tilde{\Psi}_i({\bf k}) \int_{0}^{\chi_{\rm rec}} d \chi \ \frac{\chi_{\rm rec} - \chi}{\chi \chi_{\rm rec}} D_\Psi(\chi)T(k)  e^{i \chi {\bf k} \cdot \hat{n}} \no \\
	& = -2 \int \frac{d^3k}{(2 \pi)^3} \tilde{\Psi}_i({\bf k}) \int_{0}^{\chi_{\rm rec}} d \chi \ \frac{\chi_{\rm rec} - \chi}{\chi \chi_{\rm rec}} D_\Psi(\chi)T(k)\ 4\pi \sum_{\ell m} i^\ell j_\ell(k\chi) Y^*_{\ell m}({\bf k})Y_{\ell m}({\bf \hat{n}}) \no \\
	& =  \sum_{\ell m} \left[ \int \frac{d^3k}{(2 \pi)^3} \tilde{\Psi}_i({\bf k}) Y^*_{\ell m}({\bf k}) \left(-8\pi i^\ell \int_{0}^{\chi_{\rm rec}} d \chi \ \frac{\chi_{\rm rec} - \chi}{\chi \chi_{\rm rec}} D_\Psi(\chi)T(k)j_\ell(k\chi) \right) \right]  Y_{\ell m}({\bf \hat{n}}) \no \\
	& = \sum_{\ell m} a^\phi_{\ell m} Y_{\ell m}({\bf \hat{n}}).
\end{align}
The above expression allows us to read off the lensing multipole coefficients:
\begin{equation} \label{eq:almphi}
	a^\phi_{\ell m} =  \int \frac{d^3k}{(2 \pi)^3} \tilde{\Psi}_i({\bf k}) Y^*_{\ell m}({\bf k}) \Delta_\ell^\phi(k),
\end{equation}
where the linear lensing transfer function is 
\begin{equation} \label{eq:transferphi}
	\Delta_\ell^\phi(k) = -8\pi i^\ell \int_{0}^{\chi_{\rm rec}} d \chi \ \frac{\chi_{\rm rec} - \chi}{\chi \chi_{\rm rec}} D_\Psi(\chi)T(k) j_\ell(k\chi) .
\end{equation}
The lensing power spectrum, $C_\ell^{\phi\phi}$, can be computed via the relation $\left< a^\phi_{\ell m}a^{\phi,*}_{\ell' m'} \right>= C_\ell^{\phi\phi} \delta_{\ell\ell'}\delta_{mm'} $, or equivalently, 
\begin{align}
	\left< a^\phi_{\ell m}a^{\phi,*}_{\ell' m'} \right> & = \int \frac{d^3k}{(2 \pi)^3} \int \frac{d^3k'}{(2 \pi)^3}  \left< \tilde{\Psi}_i({\bf k})\tilde{\Psi}_i({\bf k'})\right> \Delta_\ell^\phi(k) \Delta_{\ell'}^{\phi,*}(k') Y^*_{\ell m}({\bf k})Y_{\ell' m'}({\bf k'}) \no \\
	& = \int \frac{d^3k}{(2 \pi)^3}  P_\Psi(k) \Delta_\ell^\phi(k) \Delta_{\ell'}^{\phi,*}(k) Y^*_{\ell m}({\bf k})Y_{\ell' m'}({\bf k}) \no \\
	& = \int dk \frac{k^2}{(2 \pi)^3}  P_\Psi(k) \Delta_\ell^\phi(k) \Delta_\ell^\phi(k)\ \delta_{\ell\ell'}\delta_{mm'}
\end{align}
\begin{align}
	C_\ell^{\phi\phi} & = 4 \int dk \frac{2k^2}{\pi} P_\Psi(k) \int_{0}^{\chi_{\rm rec}} d \chi \ \frac{\chi_{\rm rec} - \chi}{\chi \chi_{\rm rec}} D_\Psi(\chi)T(k) j_\ell(k\chi) \no \\
	& \quad \times \int_{0}^{\chi_{\rm rec}} d \chi' \ \frac{\chi_{\rm rec} - \chi'}{\chi' \chi_{\rm rec}} D_\Psi(\chi')T(k) j_\ell(k\chi') \no \\
	& \simeq 4 \int_{0}^{\chi_{\rm rec}} \frac{d\chi}{\chi^2} P_\Psi(k) \left( \frac{\chi_{\rm rec} - \chi}{\chi \chi_{\rm rec}} \right)^2 D_\Psi(\chi)^2T(k)^2 \Big|_{k\rightarrow(\ell+1/2)/\chi} . \label{eq:Clphiphi}
\end{align}
The result is shown in figure~\ref{fig:Clphiphi}.

\bibliographystyle{JHEP}
\bibliography{psz}

\end{document}